\documentclass[12pt]{article}
\usepackage[english]{babel}
\usepackage[authoryear,numbers]{natbib}
\usepackage{lineno}
\usepackage{nohyperref}
\usepackage{amsmath,amsthm}
\usepackage{algorithm}
\usepackage{algpseudocode,algorithmicx}
\usepackage{graphicx}
\usepackage{amssymb}
\usepackage{multirow}
\usepackage{caption}
\usepackage{color,ctable}
\usepackage[mathcal]{eucal}
\usepackage{times}

\newcommand{\blind}{0}

\begin{document}

\def\spacingset#1{\renewcommand{\baselinestretch}%
{#1}\small\normalsize} \spacingset{1}


\if0\blind
{
  \title{\bf Incremental Mixture Importance Sampling with Shotgun optimization}
  \author{Biljana Jonoska Stojkova\thanks{Biljana Jonoska Stojkova is Statistician, Department of Statistics, University of British Columbia, Vancouver, BC, Canada,V6T 1Z4 (e-mail: b.stojkova@stat.ubc.ca); and David A. Campbell is Associate Professor, Department of Statistics and Actuarial Science, Simon Fraser University, Surrey, BC, Canada, V3T 0A3  (e-mail: dac5@sfu.ca).  
}\hspace{.2cm}\\
Department of Statistics,
University of British Columbia \\
     and \\
    David A. Campbell \\
    Department of Statistics and Actuarial science, Simon Fraser University}
  \maketitle
} \fi

\if1\blind
{
  \bigskip
  \bigskip
  \bigskip
  \begin{center}
    {\LARGE\bf Title}
\end{center}
  \medskip
} \fi

\bigskip
\begin{abstract}
This paper proposes a general optimization strategy, which combines results from different optimization or parameter estimation methods to overcome shortcomings of a single method.  
Shotgun optimization is developed as a framework which employs different optimization strategies, criteria, or conditional targets to enable wider likelihood exploration.  The introduced Shotgun optimization approach is embedded into an incremental mixture importance sampling algorithm to produce improved posterior samples for multimodal densities and creates robustness in cases where the likelihood and prior are in disagreement.  Despite using different optimization approaches, the samples are combined into samples from a single target posterior.   The diversity of the framework is demonstrated on parameter estimation from differential equation models employing diverse strategies including numerical solutions and approximations thereof.  Additionally the approach is demonstrated on mixtures of discrete and continuous parameters and is shown to ease estimation from synthetic likelihood models. 
R code of the implemented examples is stored in a zipped archive (codeSubmit.zip).
\end{abstract}

\noindent%

\spacingset{1.45}
\section{Introduction}

Sampling from a posterior density is challenging when the posterior modes are separated with deep valleys of  low probability   or when the posterior space is rife with many minor modes, ripples and ridges. Theoretically, standard Metropolis-Hastings or Gibbs algorithms converge to the target density if run infinitely long.  Tempering methods  such as Simulated Tempering \citep{MariParisi,ThompGeyer,zhang2008comparison} and Parallel Tempering \citep{swendsen1986replica,Geyer,Hukushima},  are random-walk variants designed to efficiently deal with sampling from multi-modal distributions. However, Parallel Tempering could exacerbate topological challenges of the posterior if the prior is inconsistent with the likelihood, trapping the sampler in a local mode \citep{SFT}.

Importance sampling algorithms such as Sampling Importance  Re-sampling (SIR) \citep{rubin1987calculation, rubin1988using,poole2000inference,alkema2011probabilistic} or Sequential Monte Carlo variants (SMC) \citep{del2006sequential} take advantage of computing the sampling weights in parallel.
The difficulty with importance sampling methods is  choosing the initial importance sampling density to cover the important modes of the target density. The prior is often chosen to be this initial importance density.

A frequentist alternative to MCMC methods would be to use optimization in order to find the modes, but in the presence of well isolated multiple modes, different starting points for the optimizer  result in multiple optima. Then the problem shifts to finding a way to combine these local optima.

Incremental Mixture Importance Sampling with Optimization (IMIS-Opt) \citep{raftery2010estimating} is designed to discover all the important posterior modes by using the prior as a starting point for optimization, and  then building a posterior through incrementally added optimized local posterior approximations. However, if the prior disagrees with the likelihood, i.e., if the prior covers the basin of attraction of local but not global likelihood modes, then the IMIS-Opt will miss the important modes. As a remedy, one can  choose a diffuse prior, but this implies that the prior should be chosen for algorithmic convenience rather than to represent expert opinion.

In this paper, we modify the IMIS-Opt algorithm by replacing the optimization step  with a general optimization strategy, which is based on the idea that no single method outperforms other methods in every problem \citep{wolpert1997no}. The proposed  multiple-method optimization strategy balances discovery of the global and the local modes by combining  results from different regions of the posterior space, corresponding to local optima found by multiple parameter estimation methods.  We refer to this  strategy as Shotgun optimization (ShOpt), and the resulting algorithm  as Incremental Mixture Importance Sampling with Shotgun optimization (IMIS-ShOpt). The IMIS-ShOpt relies on the Shotgun optimization, rather than on the prior choice. IMIS-ShOpt does not choose the prior for optimization convenience, but reaffirms its role of conveying expert opinion.

The rest of the paper is organized as follows. Section \ref{seq:optStrat} clarifies the need for  multiple optimization techniques and discusses the differences between our Shotgun optimization strategy and the multi-objective optimization. Section \ref{IMIS-Opt} gives detailed overview of the IMIS-Opt algorithm, followed by a demonstration of the IMIS-Opt getting trapped in an unimportant mode in a simple ODE model. In Section \ref{ IMIS-ShOpt} the proposed  IMIS-ShOpt algorithm is presented. Sections \ref{example} and \ref{sec:SIR_ShOpt} 
illustrate the performance of the IMIS-ShOpt  algorithm through two examples involving ODE models. The  IMIS-ShOpt via synthetic likelihood is proposed in the Section \ref{Sec:IMIS-ShOpt-SL}, and its parameter estimation performance is  illustrated using a chaotic stochastic difference equation model. Section \ref{sec:Conclusions_ShOpt} follows with concluding remarks.

\section{Shotgun optimization} \label{seq:optStrat}

Shotgun optimization is a general methodology which is directly applicable to any model type including parameter estimation in differential equation models. The ordinary differential equation (ODE) models are particularly challenging because these models exhibit likelihood topologies featuring multiple modes, ridges and ripples. Any of the existing methods for parameter estimation in ODEs might get trapped in a local mode for reasons specific to the method used. In this paper  we  demonstrate that the IMIS-ShOpt produces accurate parameter estimates  in  ODEs by combining results from different methods. Furthermore, we showcase that the IMIS-ShOpt 
can be combined with the synthetic likelihood \citep{wood2010statistical} to draw inference in models where the likelihood is intractable or costly to evaluate.

Different competitive parameter estimation methods rely on different models (such as method of moments versus maximum likelihood estimators), or different optimization methods (such as gradient, simplex or simulated annealing). In practice, one has to decide between modifying the model specification or choosing an optimization strategy where each is tuned to the specific problem. Modifying the model leads to a variant of the desired answer, while choosing an optimization  strategy requires validation if the answers are to be trusted.

For example, for inference from ODE models, strict likelihood function optimization i.e., non-linear least squares (NLS) based on the ODE solution \citep{bates1988nonlinear,seber1989f},  
discover a local optima, whereas optimization of the profile likelihood using model based data smoothing instead of the ODE solution \citep{ramsay2007parameter} will search widely for a global mode but results in higher variance estimates \citep{wu2014modeling}. Additionally, if there are multiple important modes the profile likelihood may not find them from different initializations, but NLS will find different modes with different initializations. Hence, different optimization strategies lead to different results. Then the Shotgun optimization strategy would be constructed as a combination of these two optimization methods in order to discover local and global optima \citep{berger1999integrated, walley1999upper}.

Using Shotgun optimization introduces robustness to the shortcomings of a single method.  Combining results from different optimization or parameter estimation methods ensures that posterior space has been more fully explored. The Shotgun optimization is analogous to the ensemble methods \citep{madigan1994model,hoeting1999bayesian,friedman2001elements,mendes2012ensemble, montgomery2012improving} where relative importance of the predictions are determined using a combination of models. 
Ensemble methods rely on the notion that no particular model can fully capture the data features. Hence, some models better predict certain features of the data, while producing biased predictions in some areas. The ensemble methods overcome the induced bias by combing the models together. In the Shotgun optimization, certain methods provide better estimates of the parameters than others, and combining the results from different methods overcomes the problem of the introduced bias.

The way the Shotgun optimization combines results from different competing methods is substantially different from multi-objective optimization \citep{kuhn1951proceedings,miettinen2012nonlinear}. While multi-objective optimization is designed to optimize simultaneously several objectives, the proposed Shotgun optimization strategy is a single objective optimization that combines results from multiple criteria.

\section{Incremental Mixture Importance Sampling with Optimization} \label{IMIS-Opt}

The main objective of Incremental Mixture Importance Sampling with Optimization (IMIS-Opt) \citep{raftery2010estimating} is to iteratively construct an importance sampling distribution. The initial stage of the IMIS-Opt starts by drawing $N_{0}$ samples $\boldsymbol{\Theta}_{0}=\{ \boldsymbol{\theta}_{1},..,\boldsymbol{\theta}_{N_{0}}\}$ from  the prior and then calculating their weights based on the likelihood function. In the optimization stage, the D highest-weight points are selected to sequentially initialize the optimizer, which searches for the nearest mode in the target posterior space. Then B points, drawn from the multivariate Gaussian distribution centered at the modes found by the optimizer, are added to the current importance distribution. At each iteration of the importance stage, sampling weights are calculated, and B draws from the multivariate Gaussian distribution centered at the highest-weight point are added to the current importance sampling distribution.  The weighting and sampling  steps of the importance stage are iterated  until the importance weights are reasonably uniform.   After the stopping criterion is met,  J inputs are re-sampled with replacement from  $\{ \boldsymbol{\theta}_{1},..,\boldsymbol{\theta}_{N_{K}}\}$ with weights $(w_{1},..,w_{N_{K}})^{'}$ where $K$ is the total number of particles from the importance sampling distribution. The pseudo-code of the IMIS-Opt is given in Algorithm \ref{alg:IMIS-opt_end}.

If optimization and importance sampling stages are excluded, then the algorithm becomes a Sampling Importance Re-sampling (SIR) algorithm \citep{rubin1987calculation,rubin1988using,poole2000inference,alkema2007probabilistic}. By excluding the optimization step, the algorithm becomes IMIS \citep{hesterberg1995weighted,steele2006computing}. IMIS-Opt initializes the optimizer using the D highest-weight points which makes it a powerful method for exploring the posterior space. However, the successful mixing of the IMIS-Opt depends heavily on the consistency of the information in the prior and  likelihood, and consequently, on whether or not samples from the prior cover all the important posterior modes. The implication is that the prior should be chosen for the optimization convenience rather than using the expert knowledge. 

\begin{algorithm}[] 
	\caption{IMIS-Opt}
	\textbf{Goal: Draw samples from the target distribution $P(\boldsymbol{\theta} \mid \boldsymbol{Y})$.}
	
	\textbf{Input:} Data, model, likelihood function, prior distribution, $B$ - the number of incremental points, $D$ - the number of different initial points for the optimization, $N_{0}$ - the number of the initial samples from the prior and $J$ - the number of re-sampled points, $N$ - the number of iterations.
	
	\textbf{Initial stage:}  Draw $N_{0}$ samples $\boldsymbol{\Theta}_{0}=\{ \boldsymbol{\theta_{1}},\boldsymbol{\theta_{2}},...,\boldsymbol{\theta_{N_{0}}}\}$ from the prior distribution $P(\boldsymbol{\theta})$.
	\label{alg:IMIS-opt_end}
	\begin{algorithmic}
		
		\For{$k=1:N$}	
		
		\If{k=1} 
		\State For each $\{\boldsymbol{\theta_{i}}, i=1,..,N_{0}\}$ calculate the sampling weights:	
		\begin{equation}
		w_{i}^{(1)}=\frac{P\left(\boldsymbol{Y} \mid \boldsymbol{\theta_{i}}\right)}{\sum\limits_{j=1}^{N_{0}}{P(\boldsymbol{Y} \mid \boldsymbol{\theta_{j}})}}
		\end{equation}
		
		\State \textbf{Optimization stage:}	
		\For{$d = 1: D$}
		\State Use $\boldsymbol{\theta}^{(initial)}=\underset{\boldsymbol{\theta}}{\operatorname{argmax}} \mbox{ } \boldsymbol{w}^{(1)} (\boldsymbol{\theta})$, $\boldsymbol{\theta} \in \boldsymbol{\Theta}_{d-1}$ to initialize the optimizer and get local posterior maxima $\boldsymbol{\theta}_{d}^{(Opt)}=\underset{\boldsymbol{\theta}}{\operatorname{argmax}} \mbox{ } P(\boldsymbol{\theta} \mid \boldsymbol{Y})$ along with the corresponding inverse negative Hessian $\boldsymbol{\Sigma}_{d}^{(Opt)}$.
		\State Update $\boldsymbol{\Theta}_{d}$ by excluding $\frac{N_{0}}{D}$ nearest neighbor points, $\boldsymbol{\theta}_{k} \in \boldsymbol{\Theta}_{d-1}$, that minimize the Mahalanobis distance,
		\begin{equation}
		( \boldsymbol{\theta}_{k}-\boldsymbol{\theta}_{d}^{(Opt)})^{'} (\boldsymbol{\Sigma}_{d}^{(Opt)})^{-1} (\boldsymbol{\theta}_{k}-\boldsymbol{\theta}_{d}^{(Opt)}).
		\end{equation}
		
		\State Draw B samples $\boldsymbol{\theta}_{1:B} \sim MVN(\boldsymbol{\theta}_{d}^{(Opt)},\boldsymbol{\Sigma}_{d}^{(Opt)})$; add these samples to the importance sampling distribution and
		evaluate $H_{k}=MVN(\boldsymbol{\theta}_{1:B} \mid \boldsymbol{\theta}_{d}^{(Opt)},\boldsymbol{\Sigma}_{d}^{(Opt)})$.
		\EndFor
		
		\Else
		\State \textbf{Importance sampling stage:}	
		\State For each $\{\boldsymbol{\theta_{i}}, i=1,..,N_{k}\}$ calculate weights,	
		\begin{equation}
		w_{i}^{(k)}=\frac{cP(\boldsymbol{Y} \mid \boldsymbol{\theta_{i}})P(\boldsymbol{\theta_{i}})}{ \frac{N_{0}}{N_{k}}P(\boldsymbol{\theta_{i}}) +\frac{B}{N_{k}}\sum\limits_{s=1}^{k} H_{s}(\boldsymbol{\theta_{i}}) },
		\end{equation}
		\State where $N_{k}=N_{0}+B(D+k)$ and $c=1/\sum\limits_{i=1}^{N_{k}} w_{i}^{(k)}$ is the normalizing constant.
		
		\algstore{alg:IMIS-opt_interupt}
		
	\end{algorithmic}
\end{algorithm}
\begin{algorithm}[h]	
	
	\caption*{\textbf{Algorithm \ref{alg:IMIS-opt_end}} IMIS-Opt - continued}
	
	\begin{algorithmic}
		\algrestore{alg:IMIS-opt_interupt}
        		\State  Choose the maximum weight input $\boldsymbol{\theta}_{k}$ and estimate $\boldsymbol{\Sigma}_{k}$ as the weighted covariance of B inputs with smallest Mahalanobis distance,
		\[w_p(\boldsymbol{\theta})\left(\boldsymbol{\theta}-\boldsymbol{\theta}_{k}\right)^{'}(\boldsymbol{\Sigma_{\pi}})^{-1}\left(\boldsymbol{\theta}-\boldsymbol{\theta}_{k}\right),\]
		where the weights $w_p(\boldsymbol{\theta})$ are proportional to the average of the importance weights and the uniform weights $\frac{1}{N_{k}}$, ${\boldsymbol{\Sigma}_{\pi}}$ is the covariance of the initial importance distribution.

		\State Draw B samples $\boldsymbol{\theta}_{1:B} \sim MVN(\boldsymbol{\theta}_{k},\boldsymbol{\Sigma}_{k})$; add these points to the importance sampling distribution and
		evaluate $H_{k}=MVN(\boldsymbol{\theta}_{1:B} \mid \boldsymbol{\theta}_{k},\boldsymbol{\Sigma}_{k})$.
		\EndIf 
		\If  {$\sum\limits_{1}^{N_{k}}(1-(1-w^{(k)})^{J}) \geq J(1-\exp{(-1)})$ i.e., importance sampling weights are approximately uniform}  exit for loop 
		
		\EndIf
		\EndFor
		\State \textbf{Re-sampling stage:}	
		
		\State Re-sample J points with replacement from $\{\boldsymbol{\theta}_{1},..,\boldsymbol{\theta}_{N_{k}}\}$ and weights $(w_{1},..,w_{N_{k}})^{'}$.
	\end{algorithmic}
\end{algorithm}

\section{Incremental Mixture Importance Sampling with Shotgun optimization }  \label{ IMIS-ShOpt}

The IMIS-Opt success depends heavily on the consistency between the prior and the data. If the prior is inconsistent with the likelihood then the maximum height point needed to initialize the optimizer is in the basin of attraction of the local mode that is covered by the prior. Therefore, the sampler is prevented from fully exploring the posterior space. The Incremental Mixture Importance Sampling with Shotgun optimization (IMIS-ShOpt) builds on IMIS-Opt, by altering the optimization stage to incorporate the Shotgun optimization strategy, which consists of Q different competitive parameter estimation methods or optimization strategies. This implies using a variety of optimization methods in parallel or using a fixed optimizer on a variant of the function to optimize, such as likelihood or other objective function within the estimating framework. The Shotgun optimization strategy sequentially initializes 
Q different optimization methods (which could be run in parallel) for each of the D maximum weight points from the prior. Pseudo-code of the proposed Shotgun optimization strategy is given in Algorithm \ref{alg:IMIS-ShOpt}. Replacing the optimization step in the Algorithm 	\ref{alg:IMIS-opt_end} with the Shotgun optimization in the Algorithm \ref{alg:IMIS-ShOpt} gives the pseudo code of the IMIS-ShOpt algorithm.
\begin{algorithm}[h]  
	\caption{The Shotgun optimization}
	\begin{algorithmic}
		\State \textbf{Optimization stage:}	
		\For{$d = 1: D$}
		
		\State Find the d-th maximum weight point
		$\boldsymbol{\theta}_{d}^{(initial)}=\underset{\boldsymbol{\theta}}{\operatorname{argmax}} \mbox{ } \boldsymbol{w}^{(k)} (\boldsymbol{\theta})$, $\boldsymbol{\theta} \in \boldsymbol{\Theta}_{d-1}$ to initialize $Q$ optimizers. 
		
		\For{$q = 1: Q$}
		\State  Use q-th optimization method initialized at $\boldsymbol{\theta}_{d}^{(initial)}$ to obtain local  maxima $\boldsymbol{\theta}_{d,q}^{(Opt)}$ along with the corresponding inverse negative Hessian $\boldsymbol{\Sigma}_{d,q}^{(Opt)}$ (this step can be parallelized).
		\State Update $\boldsymbol{\Theta}_{d}$ by excluding $\frac{N_{0}}{QD}$ nearest neighbor points, $\boldsymbol{\theta}_{k} \in \boldsymbol{\Theta}_{d-1}$, that minimize the Mahalanobis distance,
		\begin{equation}
		( \boldsymbol{\theta}_{k}-\boldsymbol{\theta}_{d,q}^{(Opt)})^{'} (\boldsymbol{\Sigma}_{d,q}^{(Opt)})^{-1} (\boldsymbol{\theta}_{k}-\boldsymbol{\theta}_{d,q}^{(Opt)} ). \end{equation}
		\State Draw B samples $\boldsymbol{\theta}_{1:B} \sim MVN(\boldsymbol{\theta}_{d,q}^{(Opt)},\boldsymbol{\Sigma}_{d,q}^{(Opt)})$; add these points to the importance sampling distribution and
		evaluate $H_{k}=MVN(\boldsymbol{\theta}_{1:B} \mid \boldsymbol{\theta}_{d,q}^{(Opt)},\boldsymbol{\Sigma}_{d,q}^{(Opt)})$.
		\EndFor
		\EndFor
	\end{algorithmic}
	\label{alg:IMIS-ShOpt}
\end{algorithm}

 IMIS-ShOpt explores modes and merges the samples from different regions of the target posterior, $P(\boldsymbol{\theta} \mid \boldsymbol{Y})$, explored by the variety of criteria. Although the IMIS-ShOpt draws samples from the target posterior distribution  $P(\boldsymbol{\theta} \mid \boldsymbol{Y})$, the optimization step uses different strategies of modifying the target  posterior  to improve the exploration of the parameter space. The modification of the posterior depends on the parameter estimation method used. For example, if a parameter of interest is a location parameter, the Multiple-method optimization in IMIS-ShOpt could be comprised of Q=2  methods: the Maximum Likelihood method and the Method of Moments. Therefore, the posterior modifications targeted by different optimization methods may give different results due to the differences in topology of the posterior space.  

Sampling weights in the IMIS-ShOpt are obtained with respect to the target posterior distribution, ensuring that the unlikely points are not re-sampled in the final  stage. Hence, keeping the unlikely points in the importance sampling distribution does not harm the algorithm, but it does improve the posterior exploration.

\section{Ordinary differential equation models}
Ordinary differential equation (ODE) models are mechanistic models which describe the rate of change of system states $\boldsymbol{X}(\boldsymbol{\theta},t)$ which are realizations of a S-dimensional process $\boldsymbol{X}$ at time t with parameters $\boldsymbol{\theta} \in \boldsymbol{\Theta}^{P}$,
\begin{equation}
\frac{d\boldsymbol{X}\left(\boldsymbol{\theta},t\right)}{dt} = f\left(\boldsymbol{X}(\boldsymbol{\theta},t), \boldsymbol{\theta}\right). \label{eq:eqODE}
\end{equation}
The s-th system state,
\begin{equation}
\frac{dX_{s}\left(\boldsymbol{\theta},t\right)}{dt} = f_{s}\left(\boldsymbol{X}(\boldsymbol{\theta},t), \boldsymbol{\theta}\right),   \label{eq:eqODE_s}
\end{equation}
relies on a known function $f_{s}$ that depends on the entire set of S system states. The ODE systems are designed to capture complex phenomena using few parameters while preserving interpretability. The goal is to estimate the parameters $\boldsymbol{\theta}$, given the noisy observations  $\boldsymbol{Y}=\{y_{sj}\}$ at times $\boldsymbol{t}=\{t_{sj}\}$, for $s=1,..,S,j=1,..n_{s}$. Usually the analytical solution to  (\ref{eq:eqODE}) does not exist, and hence, a numerical solver must be used with initial state $\boldsymbol{X}(0)=\boldsymbol{X}(\boldsymbol{\theta},0)$ to obtain the solution $\boldsymbol{X}(\boldsymbol{\theta},t)$. In practice, the initial state vector is not known, and has to be estimated together with the unknown parameters $\boldsymbol{\theta}$. 

Using a Gaussian error structure centered at the solution to the ODE model in (\ref{eq:eqODE}), $\boldsymbol{X}(\boldsymbol{\theta},t)$, the likelihood for observation vector $\boldsymbol{y}_{s}=(y_{s1},..,y_{sn_{s}})^{'}$ from states is: 
\begin{equation}
P(y_{sj} \mid \boldsymbol{X}(\boldsymbol{\theta},t_{sj}), \boldsymbol{\theta})= N\left(\boldsymbol{X}(\boldsymbol{\theta}, t_{sj}), \boldsymbol{\sigma}^2_{\boldsymbol{y}_{s}}\right). \label{eq:measure}
\end{equation}
Small changes in parameters can lead to big changes in the dynamics of the model. Consequently, multi-modality, ridges and deep valleys of low-probability areas are common characteristics of the likelihoods in ODE models \citep{SFT}.
Standard random walk MCMC algorithms could easily get trapped in a local mode. Model relaxation methods that use  model based smoothing, rather than numerically solving the ODE system in (\ref{eq:eqODE}) have been designed to overcome the topological challenges \citep{ramsay2007parameter,brunel2008parameter,liang2008parameter}. These methods will be discussed in the Section \ref{ShotGunFhNODE}.

\subsection{Motivating example -- the FitzHugh-Nagumo ODE model} \label{sec:FHN_intro}
The FitzHugh-Nagumo model \citep{fitzhugh1961impulses,nagumo1962active} captures the behavior of spike potentials in the giant axon of squid neurons. The FitzHugh-Nagumo (FhN) model is described by a system of two  non-linear differential equations, corresponding to the two state variables: voltage across the membrane, $V$, and outward currents (recovery), $R$, with a vector of parameters of interest  $\boldsymbol{\theta}=\left(a, b, c\right)^{'}$,
\begin{eqnarray}
\frac{dV}{dt}  = c\left(V(t)-V(t)^{3}/3+R(t)\right) \mbox{ and }
\frac{dR}{dt}  = -\frac{1}{c}\left(V(t)-a+bR(t)\right).
\label{eq:FHN-ODE}
\end{eqnarray} 
The analytic solution of the ODE system (\ref{eq:FHN-ODE}) does not exist and therefore the numerical solution to the system can be used with initial states values $\{V(0),R(0)\}=\{V(\boldsymbol{\theta},0), R(\boldsymbol{\theta},0)\}$. The likelihood follows the measurement error model in (\ref{eq:measure}), centered about the solution of (\ref{eq:FHN-ODE}), $V(\boldsymbol{\theta},t)$ and $R(\boldsymbol{\theta},t)$, 
\begin{eqnarray}
\boldsymbol{Y}_{V}(t) \mid \boldsymbol{\theta} \sim N \left(V(\boldsymbol{\theta},t),\sigma^{2}_{V}\right) \mbox{ and }
\boldsymbol{Y}_{R}(t) \mid \boldsymbol{\theta} \sim  N\left(R(\boldsymbol{\theta},t),\sigma^{2}_{R}\right).  
\label{eq:FHN-ODE-likelihood}
\end{eqnarray} 
The  vector of parameters of interest in the  model including the initial points is 

$\boldsymbol{\theta}=\left(a, b, c,\sigma^{2}_{V}, \sigma^{2}_{R}, V(0), R(0)\right)^{'}$. For expositional simplicity, we consider a one parameter model while holding the rest of the parameters fixed to the values, 

$\left(a=0.2, b=0.2, \sigma^{2}_{V}=0.05^2, \sigma^{2}_{R}=0.05^2, V(0)=-1, R(0)=1\right)^{'}$, with $\boldsymbol{\theta}=c$ being the only parameter to estimate.

As an illustrative example, placing a prior which assumes that oscillations occur an integer multiples of the true frequency of the oscillation, induces inconsistency between the prior and the data. 
For example, the prior,
\begin{eqnarray}
P(c)=N \left(14,2 \right),  \label{eq:prior-c}  
\end{eqnarray}
suggests that there is only one full oscillation in the system (Figure~\ref{fig:ResampTraj} A), while for the true value $c=3$, the data exhibit two full oscillations (Figures~\ref{fig:ResampTraj} B and C).

\begin{figure}[!h]
	\centerline{\includegraphics[scale=0.25]{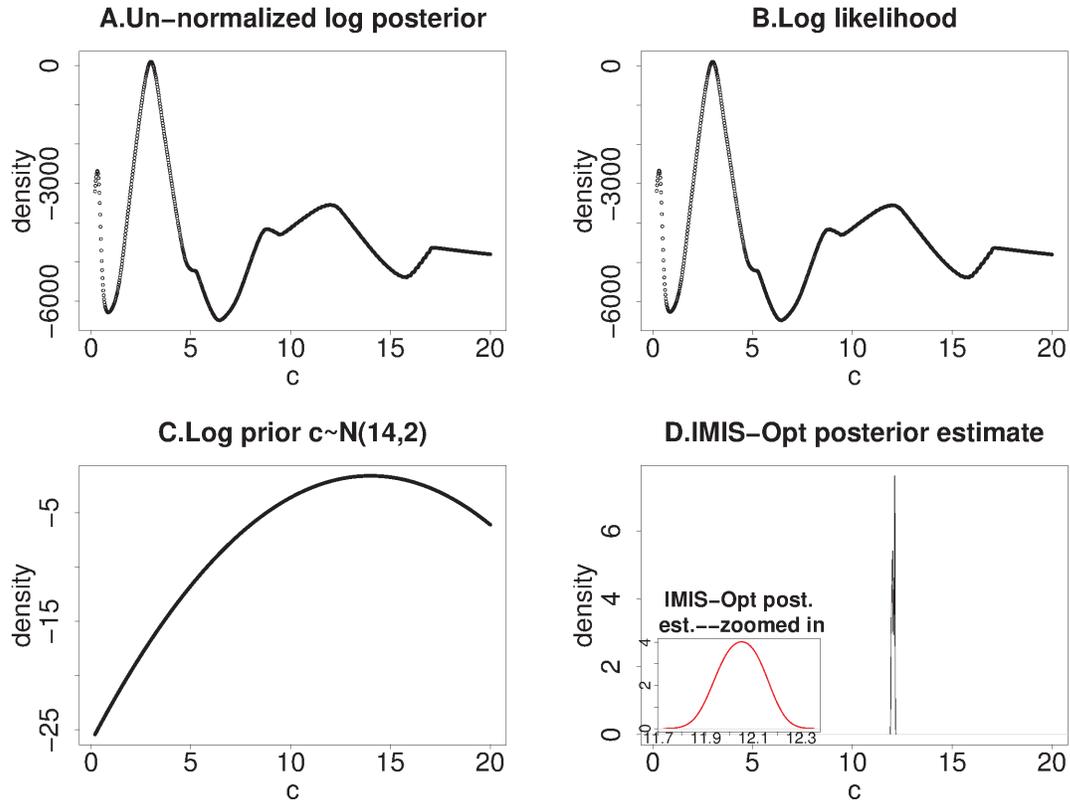}}
	\caption{The FhN-ODE model -- impact of the disagreement between the log-likelihood and log posterior (plots A and B) and log prior (plot C) on the IMIS-Opt posterior estimate (plot D). The IMIS-Opt was run with D=3, B=1000 and J=10000.}
	\label{fig:FHN-llik}
\end{figure}

Figure~ \ref{fig:FHN-llik} A and B show that the likelihood and the target distribution exhibit multiple modes separated with deep valleys of near-zero probability regions measuring several thousands on the log scale. Standard random walk algorithms could get easily trapped in an unimportant local mode around $c=12.05$. 
The prior given by the equation (\ref{eq:prior-c}) covers only one of the  local modes in the likelihood (Figures~\ref{fig:FHN-llik} A and C), and does not cover the basin of attraction of the global mode.
Consequently, IMIS-Opt is trapped in the local mode that is covered by the prior (Figure \ref{fig:FHN-llik}D). 

\section{Illustrative example -- the FitzHugh-Nagumo model revisited} \label{example}

We illustrate the performance of the IMIS-ShOpt using the one parameter FhN-ODE model from Section \ref{sec:FHN_intro} (Model 1) and the full FhN-ODE model (Model 2) with $\boldsymbol{\theta}=\left(a, b, c,\sigma^{2}_{V}, \sigma^{2}_{R}, V(0), R(0)\right)^{'}$. For comparison, the results from the performance of the IMIS-Opt on the Model 1 are also presented and discussed. Table \ref{tbl:models} presents prior specifications of the two models.

\begin{table}[h]
	\caption{The two FhN models -- prior specifications}
	\centering
	\resizebox{\textwidth}{!}{\begin{tabular}{rlllllll}
			\hline
			&  a & b & c & $\sigma^{2}_{V}$ & $\sigma^{2}_{R}$ &  $V(0)$ & $R(0)$  \\ 
			\hline
			Model 1  &  0.2 & 0.2 & $N\left(14,2 \right)$ & 0.05 & 0.05 & -1 &  1 \\ 
			Model 2  &   $N\left(0,.4 \right)$ & $N\left(0,.4 \right)$ & $N\left(14,2\right)$ & $IGamma\left(3,3\right)$ & $IGamma \left(3,3\right)$ &  $N\left(-1,.5\right)$ & $N\left(1,.5\right)$   \\ 
			\hline
	\end{tabular}}
	\caption*{The two FhN models -- in the Model 1, prior has been assigned only for the parameter $c$, while the rest of the parameters are fixed to their true values. In Model 2 prior distributions have been  assigned for all parameters.}
	\label{tbl:models}
\end{table}

\subsection{Shotgun optimization strategy for the FhN model}  \label{ShotGunFhNODE}

The Shotgun optimization strategy used to estimate the parameters of the FhN model comprises three different parameter estimation methods in ODE models: $i).$  Non-linear Least Squares (NLS) \citep{bates1988nonlinear,seber1989f}, $ii).$ Two Stage estimator \citep{varah1982spline, brunel2008parameter,liang2008parameter} and $iii).$ Generalized Profiling (GP) \citep{ramsay2007parameter}. All three are described bellow.

\paragraph*{The NLS method}  \label{sec:NLS}

Following \cite{bates1988nonlinear}, the maximum likelihood estimate $\boldsymbol{\hat{\theta}}$ is obtained by minimizing the negative log-likelihood, which in Gaussian distribution (as per (\ref{eq:FHN-ODE-likelihood})) becomes a sum of squared difference between observations and the numerical solution solution to the ODE model in (\ref{eq:eqODE}), 

\begin{equation}
\hat{\boldsymbol{\theta}}= \arg\min_{\boldsymbol{\theta}} \sum\limits_{s=1}^{S}
\sum\limits_{j=1}^{n_{s}} \left[y_{sj}-\boldsymbol{X}(\boldsymbol{\theta},t_{sj})\right]^{2}. \label{eq:NLSOptim}
\end{equation}
The NLS method has several drawbacks. First, in order to minimize the  sum of squared error in (\ref{eq:NLSOptim}), NLS requires numerically solving the ODE system in (\ref{eq:eqODE}) at each evaluation of the optimization criteria, which, in turn, requires the initial system states. NLS  estimates depend on the initial guesses of the parameters of interest especially in the cases when the sum of square error function in (\ref{eq:NLSOptim}) exhibits multiple modes. As a result, the starting points determine  whether the parameter estimate will converge to a local or global mode. Consequently, the NLS performs well in the cases when the neighborhood of the true parameters values are used as initial optimization guesses.

\paragraph*{The Two-Stage method}  \label{sec:Two-Stage}

The Two-Stage method first smooths the data as an estimate  $\hat{\boldsymbol{X}}(\boldsymbol{\theta},t)$ and then differentiates that smooth to approximate $\frac{d\boldsymbol{X}(\boldsymbol{\theta},t)}{dt}$ \citep{varah1982spline,  brunel2008parameter, liang2008parameter}. Parameter estimates are obtained by maximizing fidelity to the ODE model in (\ref{eq:eqODE}) using the estimates from the smoothing step.

The local polynomial procedure \citep{fan1996local} 
approximates the $s$-th state $\boldsymbol{X}_{s}(\boldsymbol{\theta},t_{sj})$ by a $\nu$-th order polynomial, in a neighborhood of the time point $t_{s0}$, with $a_{i}(\boldsymbol{\theta},t_{s0})=\boldsymbol{X}_{s}^{(i)}(\boldsymbol{\theta},t_{s0})$ for $i=0,..,\nu$,
\begin{eqnarray}
\boldsymbol{X}_{s}(\boldsymbol{\theta},t_{sj})  &\approx&  \boldsymbol{X}_{s}(\boldsymbol{\theta},t_{s0})+(t_{sj}-t_{s0})\boldsymbol{X}_{s}^{(1)}(\boldsymbol{\theta},t_{s0})+..+(t_{sj}-t_{s0})^{s}\boldsymbol{X}_{s}^{(\nu)}(\boldsymbol{\theta},t_{s0})/ \nu! \nonumber \\
&=&
\sum\limits_{i=0}^{\nu}  a_{i}(\boldsymbol{\theta},t_{s0})(t_{sj}-t_{s0})^{i},\mbox{ for } s=1,..,S,j=1,..,n_{s}.
\end{eqnarray}

Following \cite{fan1996local}, the estimators
$\widehat{\boldsymbol{X}^{(i)}_{s}}(\boldsymbol{\theta},t), i=0,1,$ are obtained by minimizing the locally weighted least-square criterion,
\begin{equation}
\sum\limits_{j=1}^{n_{s}}\left[ y_{sj}-\sum\limits_{i=0}^{\nu}  a_{i}(t_{sj}-t_{s0})^{i}\right]^{2}K_{h}(t_{sj}-t_{s0}), \label{eq:localPol}
\end{equation}
where h controls the size of the neighborhood around $t_{s0}$, $K_{h}(.)=K_{h}/h$ controls the weights, and $K(.)$ is a Kernel weight function.

In the second stage, the estimate $\hat{\boldsymbol{\theta}}$ is  obtained by minimizing the sum of squared difference between the derivative estimate and the derivative from the ODE model,
\begin{equation}
\hat{\boldsymbol{\theta}}=\arg\min_{\boldsymbol{\theta}}\sum\limits_{s=1}^{S}\sum\limits_{j=1}^{n_{s}} \left[ \widehat{\boldsymbol{X}^{(1)}_{s}}(\boldsymbol{\theta},t_{sj}) - f_{s}(\hat{\boldsymbol{X}}(\boldsymbol{\theta},t_{sj}),\boldsymbol{\theta} )  \right]^{2}. \label{eq:TwoStage}
\end{equation}

Although the objective function (\ref{eq:TwoStage}) resembles the least squares, the error term is not independently distributed. Hence, the estimator $\hat{\boldsymbol{\theta}}$ is called pseudo-least squares (PsLS) estimator. Alternatively, the SIMEX  \citep{carroll2006measurement} algorithm can be used to deal with measurement error in covariates for nonlinear regression models.

The Two-Stage method is computationally more efficient than the NLS, since it avoids employing the numerical solver at each evaluation of the objective function. However, this gain of computational efficiency comes at the cost of accuracy. Namely, in the first stage the data are smoothed without using the ODE model information. The ODE model is only used in the second stage to obtain $\hat{\boldsymbol{\theta}}$ based on the first stage smoothing results. Separating the estimation procedure in two stages results in a reduced estimation accuracy of the ODE parameters \citep{ding2014estimation}. Combining the Two-Stage and the NLS method can improve parameter estimates by first obtaining the neighborhood of the estimates from the Two-Stage method and then using them as initial points for the NLS \citep{wu2014modeling}.

\paragraph*{The Generalized Profiling method}   \label{sec:GP}

Avoiding the numerical solution to the ODE system, the  GP method uses collocation to smooth out the data which is governed by the ODE model through penalizing the deviation at the level of the derivative.

The GP is a parameter cascade optimization procedure which first profiles out the basis coefficients $\boldsymbol{c}$ for basis functions $\boldsymbol{\Phi}(t)$ of the ODE model based data smooth, and then  estimates the ODE parameters using the profile likelihood.

The model based data smoothing is performed to obtain the basis functions coefficients. Being nuisance parameters, the basis coefficients are obtained by keeping $\boldsymbol{\theta}$ fixed, while optimizing the inner criterion, 
\begin{equation}
G\left(\boldsymbol{C} \mid \boldsymbol{\theta}, \boldsymbol{\lambda}, \boldsymbol{Y}\right)=\sum\limits_{s=1}^{S_{0}} \sum\limits_{j=1}^{n_{s}}  \omega_{sj} \left[y_{sj}- \boldsymbol{\Phi}_{s}(t_{sj}) \boldsymbol{c_{s}}\right]^2 + \sum\limits_{s=1}^S \lambda_{s} \int\limits_{\boldsymbol{T}} \left[\frac{d  \boldsymbol{\Phi}_{s}(t)\boldsymbol{c_{s}}}{dt}-f_{s}(\boldsymbol{\Phi}(t) \boldsymbol{c}, \boldsymbol{\theta})\right]^{2}dt, \label{eq:innerOpt}
\end{equation}
where $t$ is integrated over the interval of observation times and $S_{0}$ is the dimension of the observed system states such that $S_{0} \leq S$.
The first term of G represents a weighted sum of squares which is a measure of how well the observed states are approximated by the basis functions, while the second term of G measures the fidelity of the basis functions to the ODE model. The smoothing parameter $\lambda$ controls the trade-off between fit to the data and fidelity to the ODE model.  For notational simplicity, the dependence of  $\boldsymbol{c}_{s}$ on $\boldsymbol{\theta}$ in (\ref{eq:innerOpt}) is omitted. Hence, having $\boldsymbol{c}_{s}(\boldsymbol{\theta})$ in (\ref{eq:innerOpt}) implicates that for any set of $\boldsymbol{\theta}$ the inner optimization criteria  is optimized with respect to  the basis functions coefficients $\boldsymbol{c}_{s}$.  

The outer optimization criterion,  
\begin{equation}
J\left(\boldsymbol{\theta} \mid \boldsymbol{C},\boldsymbol{Y}\right)=\sum\limits_{s=1}^{S_{0}} \sum\limits_{j=1}^{n_{s}} \omega_{sj} \left[y_{sj}-\boldsymbol{\Phi}_{s}(t_{sj})\boldsymbol{c}_{s}(\boldsymbol{\theta})\right]^2,   \label{eq:outerOpt}
\end{equation}
produces  $\hat{\boldsymbol{\theta}}$ estimates using the  basis functions coefficients estimates obtained from the inner optimization.

\subsection{Performance of the Shotgun optimization strategy in the FhN-ODE model.}

The target posterior is based on the likelihood in (\ref{eq:NLSOptim}). Rather than optimizing the posterior target distribution to find the important modes as per the IMIS-Opt optimization step, the Shotgun optimization strategy in the IMIS-ShOpt employs: the NLS method in (\ref{eq:NLSOptim}), the Two-Stage method  in (\ref{eq:TwoStage}), and the GP parameter cascade optimization  in (\ref{eq:innerOpt}) and (\ref{eq:outerOpt}). In the Shotgun optimization, the parameter estimates $\hat{\boldsymbol{\theta}}$ are obtained by combining the results from different optimization criteria, while the Hessian matrices evaluated at  $\hat{\boldsymbol{\theta}}$ are obtained  using the target posterior. 

The three methods (NLS, Two-Stage and GP) combined together discovered global and local optima. The prior of the parameter $c$ covers only the unimportant local mode of the target posterior centered around $c$=12.05, and therefore, the initial particles in the IMIS-ShOpt are in the basin of attraction of that local mode, thus missing the global mode. The results from the NLS were highly affected by the initial points, and consequently, the optima from the NLS were in the basin of attraction of the local mode at $c=12.05$. The GP method was occasionally discovering both the local and the global mode. The two-stage method proved to be the least sensitive to the initial points and hence, it was the only method among the three that discovered the global mode with any starting point. The exploration of global and local maxima obtained from the Shotgun optimization is the goal of  IMIS-ShOpt.

Shotgun optimization strategy is computationally efficient due to its ability to run in parallel its constituting methods (here NLS, Two-Stage and GP). 
Table~\ref{tbl:CompTime} shows the computational time in seconds needed to run the IMIS-ShOpt for the Model 1 and Model 2 and the IMIS-Opt for the Model 1. The IMIS-ShOpt in Model 1 is faster than IMIS-Opt for the Model 1, because the Shotgun optimization explores the posterior space efficiently thus enabling the sampler to converge in just 2 iterations. By contrast, the optimization stage in the IMIS-Opt is less efficient and it takes 150 iterations for the algorithm to converge.

\begin{table}[!h]
	\caption{Computational time} \label{tbl:CompTime}
	\centering
	\begin{tabular}{rrrr}
		\hline
		& IMIS-Opt, Model 1 & IMIS-ShOpt, Model 1 & IMIS-ShOpt, Model 2 \\ 
		\hline
		& 697.325 & 521.531 & 3784.042 \\ 
		\hline
	\end{tabular}
	\caption*{Wall-clock time in seconds of the runs from IMIS-ShOpt and IMIS-Opt on Model 1 and Model 2.}   
\end{table}

\subsection{Results}
\begin{figure}[h!]
	\vspace{.000003in}
	\centering
    {\begin{tabular}{@{}cc@{}}
     \includegraphics[scale=0.14]{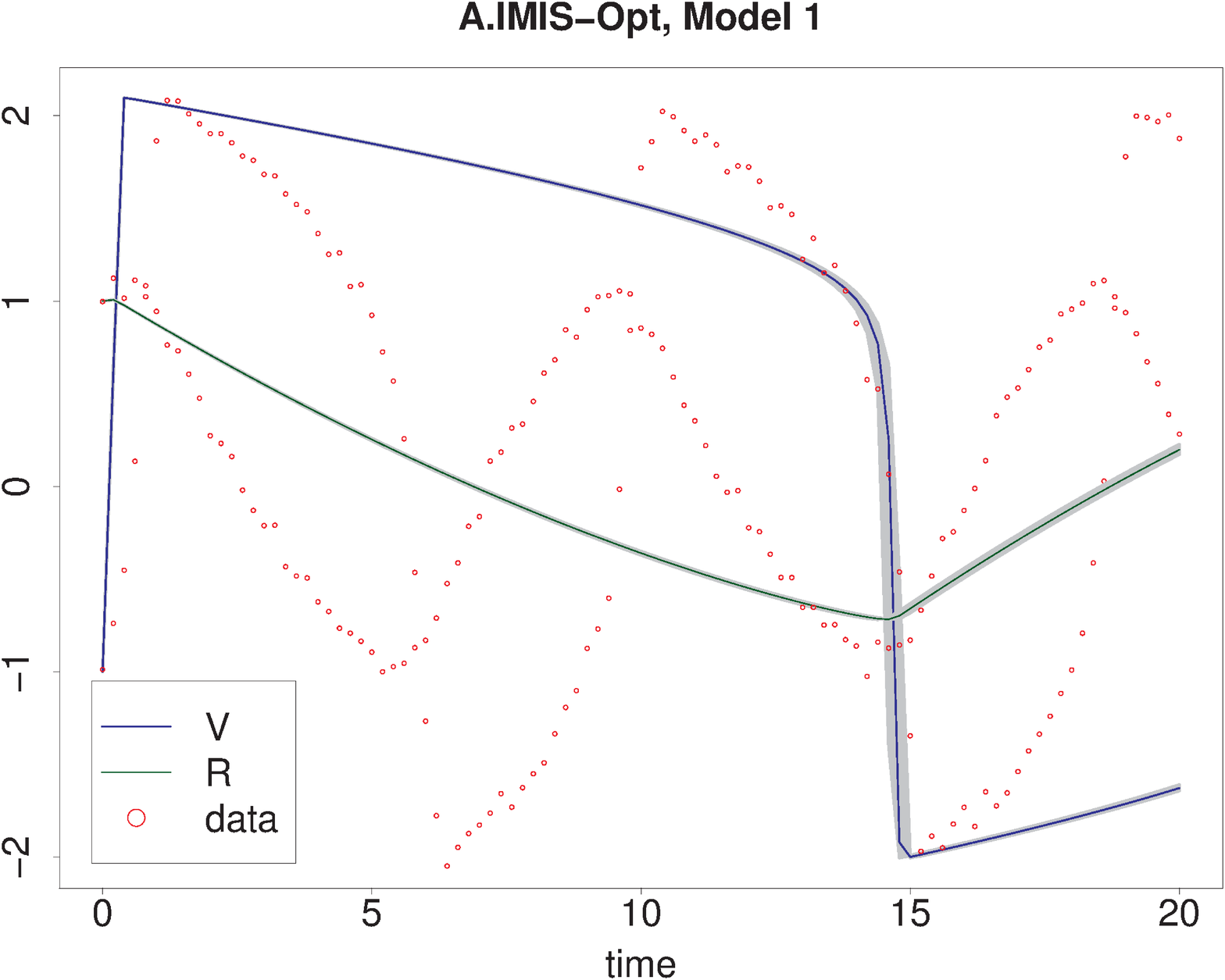}&
     \includegraphics[scale=0.14]{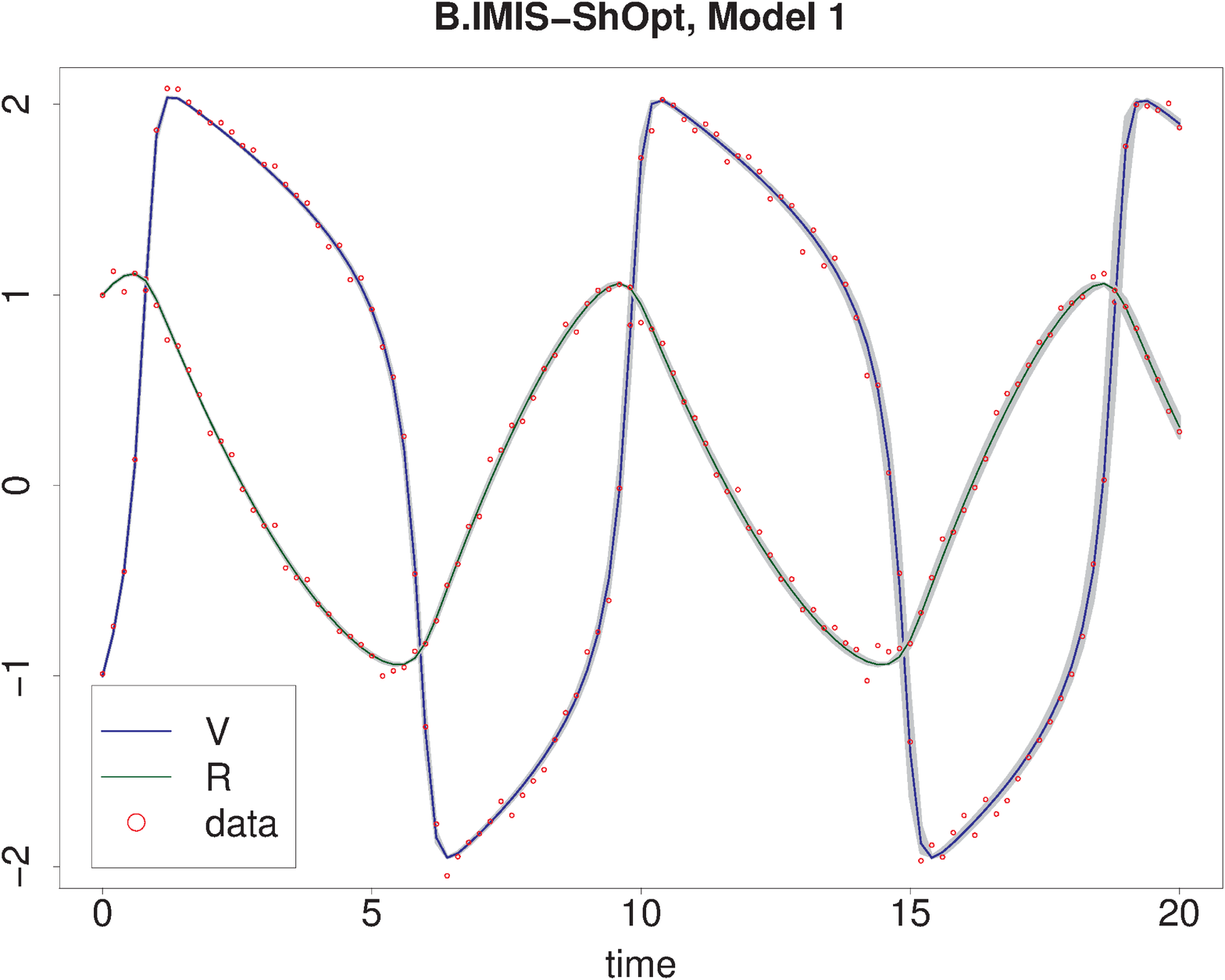}
     \end{tabular}\par}
     \vspace{0.3in}
    {\begin{tabular}{@{}r@{}}
     \includegraphics[scale=0.14]{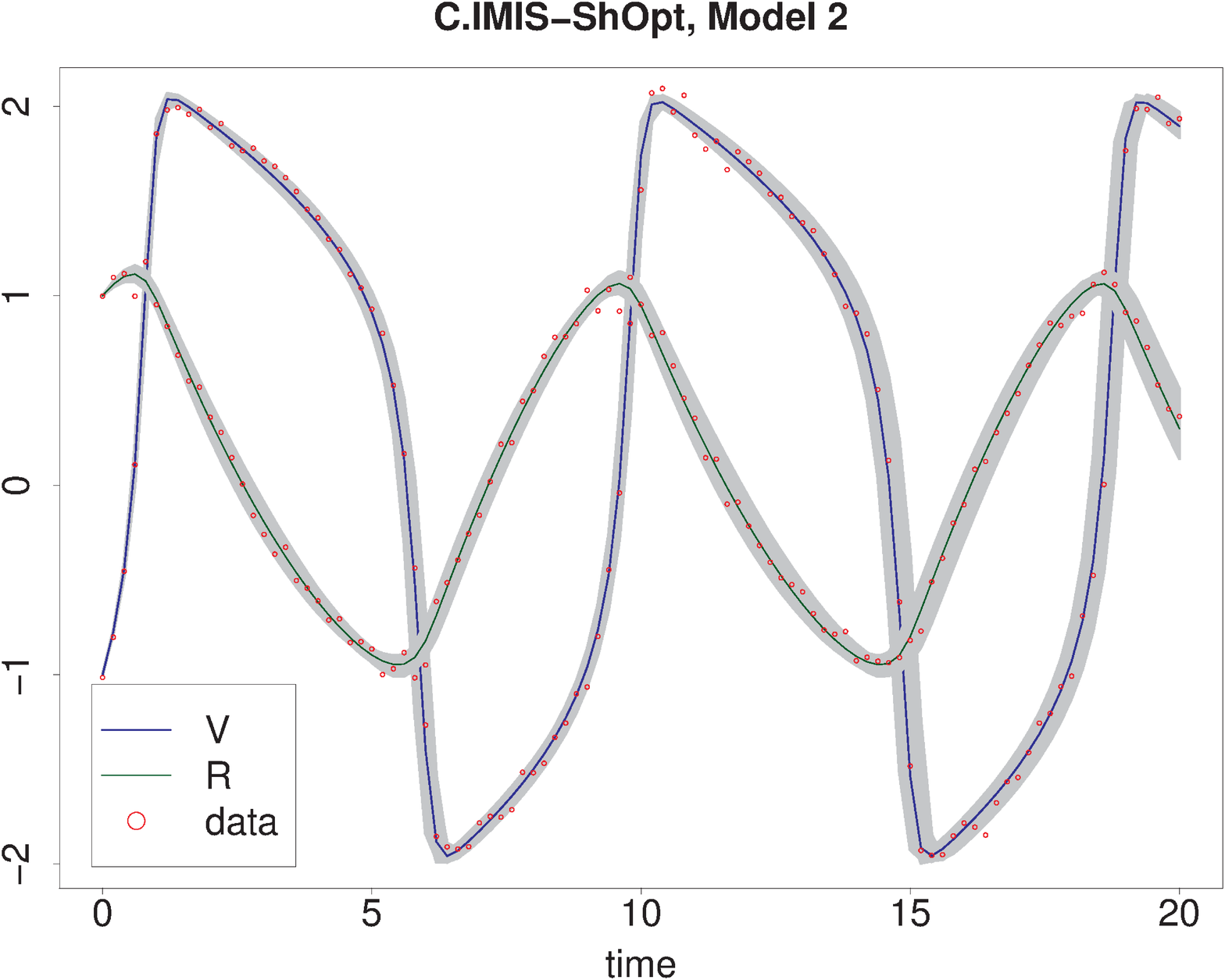} 
     \end{tabular} \par}
	\vspace{.000003in}
	\caption{The FhN-ODE model -- re-sampled trajectories using IMIS-Opt on Model 1 (plot A), IMIS-ShOpt on Model 1 (plot B) and IMIS-ShOpt  on Model 2 (plot C). The gray lines represent 10000 re-sampled trajectories, the solid thick blue and green thin lines correspond to the re-sampled trajectories at the posterior mean values for the state variables V and R, respectively. The red points represent the data, which were simulated from the vector of true parameters values $\boldsymbol{\theta}=(a=0.2,b=0.2,c=3,V(0)=-1,R(0)=1)^{'}$. The IMIS-Opt was run with D=3, B=1000 and J=10000. The IMIS-ShOpt for both models, Model 1 and Model 2, was run with D=30, Q=3, B=1000 and J=10000.}
	\label{fig:ResampTraj}
\end{figure}

Figure~\ref{fig:ResampTraj}, B and C demonstrate that although the prior for the parameter $c$ does not cover the global mode, the IMIS-ShOpt recovers the two and a half oscillations of the true trajectories in Model 1 and Model 2. By contrast, the re-sampled trajectories obtained from the IMIS-Opt (Figure~\ref{fig:ResampTraj} A), recover only one oscillation of the true trajectories, while missing the other one-and-a-half oscillation. If IMIS-Opt used a Stochastic global optimizer or an evolutionary optimizer instead of gradient descent, the global maximum could have been found.

\section{Illustrative example -- Susceptible-Infected-Removed (SIR) epidemiological model} 
\label{sec:SIR_ShOpt}

In this section we consider a Susceptible-Infected-Removed (SIR) epidemiological model using the data from the second black plague outbreak in the village of Eyam UK, from June 19, 1666 to November 1, 1666 \citep{massad2004eyam}. Since the village had been quarantined, the population size is fixed to N=261 and is stratified into states of susceptible S(t), infected  I(t) and removed R(t) individuals, N=S(t)+I(t)+R(t). $R(t)$ corresponds to the number of deaths up to time t, because there is no recovery from the plague \citep{campbell2014anova,golchi2016sequentially}.
The following system of ordinary differential equations (ODE) models the disease spread dynamics:
\begin{align}
\label{eq:SIR_ShOpt}
\frac{dS}{dt}   =-\beta S(t)I(t), \mbox{ }  \frac{dI}{dt}   = \beta S(t)I(t)-\alpha I(t), \mbox{ } \frac{dR}{dt}    = \alpha I(t)
\end{align}
where $\alpha$  describes the rate of death once the individual is infected and $\beta$ describes the plague transmission. In order for the ODE system in (\ref{eq:SIR_ShOpt}) to be numerically solved, the initial states $S(0),I(0)$ and $R(0)$ are required. Since the number of removed at the initial time is 0, $R(0)=0$, it follows that $S(0)=N-I(0)$, the initial states of the system reduce to $I(0)$. Hence, parameters of the model are $\boldsymbol{\theta}=\left(\alpha,\beta,I(0)\right)^{'}$. The data $\boldsymbol{Y}= (y_{1},.., y_{n})^{'}$ comprise of the cumulative number of deaths up to times $(t_{1},..,t_{n}), n=136$. The likelihood of the data followed a binomial distribution with expected value equal to the solution  $R_{(\alpha,\beta,I(0))}(t)$ to the system in (\ref{eq:SIR_ShOpt}). 

The states  S(t) and I(t) are not observed, however, the number of infected at the end of the plague is 0,  and the number of infected at time one before the end of the plague must therefore equal 1 \citep{campbell2014anova}. Two additional data points on number of infected individuals $\boldsymbol{X}=(x_{n-1}=1,x_{n}=0)^{'}$ at times $(t_{n-1},t_{n})^{'}$ were modeled using binomial distribution with expected value equal to the solution $I_{(\alpha,\beta,I(0))}(t)$ to the system in (\ref{eq:SIR_ShOpt}) at $t \in ( t_{n-1},t_{n})^{'}$ time points,
\begin{eqnarray}
\label{eq:llik_ShOpt}
P( \boldsymbol{Y} \mid \alpha,\beta,I(0) ) & = & \prod_{i=1}^{n}    \mbox{Binomial} \bigg(y_{i} \mid N, \frac{R_{(\alpha,\beta,I(0))}(t_{i})}{N} \bigg) \times \nonumber \\  
& &\prod_{i=n-1}^{n} \mbox{Binomial}  \bigg(x_{i} \mid  N, \frac{I_{(\alpha,\beta,I(0))}(t_{i})}{N} \bigg). 
\end{eqnarray}
Prior distributions for $\boldsymbol{\theta}=(\alpha,\beta,I(0) )^{'}$ were chosen to be:
\begin{equation}
\alpha,\beta \sim  \mbox{Gamma}(1,1), 
I(0) \sim   \mbox{Binomial}(N,\frac{5}{N}). \label{eq:priors}
\end{equation}

\subsection{Shotgun optimization strategy used in SIR-ODE model}

The challenge of this model is the mixture of discrete and continuous parameters.  Consequently we employ the Shotgun optimization strategy targeting different conditional likelihoods rather than different optimization algorithms.  Shotgun optimization applied to the SIR-ODE model uses the D=3 highest weights points to initialize the optimizer, and Q=10 likelihoods conditional on fixed discrete values of $I(0) \in \{1,2,3,..,10\}$. 
The Hessian matrix was evaluated using (\ref{eq:llik_ShOpt}). 
Implementation details are given in the Appendix \ref{app:ImplemIMIS-ShOpt}.

Table~\ref{tbl:CompTimeSIR} shows the computational time in seconds needed to run the IMIS-ShOpt in comparison to that of the IMIS-Opt for the SIR model.

\begin{table}[!h]
	\caption{Computational time} \label{tbl:CompTimeSIR}
	\centering
	\begin{tabular}{rrr}
		\hline
		& IMIS-Opt & IMIS-ShOpt \\ 
		\hline
		& 169.244 & 319.739 \\ 
		\hline
	\end{tabular}
	\caption*{Wall-clock time in seconds of the runs from IMIS-ShOpt and IMIS-Opt on the SIR model.}   
\end{table}

\begin{figure}[h]
	\centerline{\includegraphics[scale=0.26]{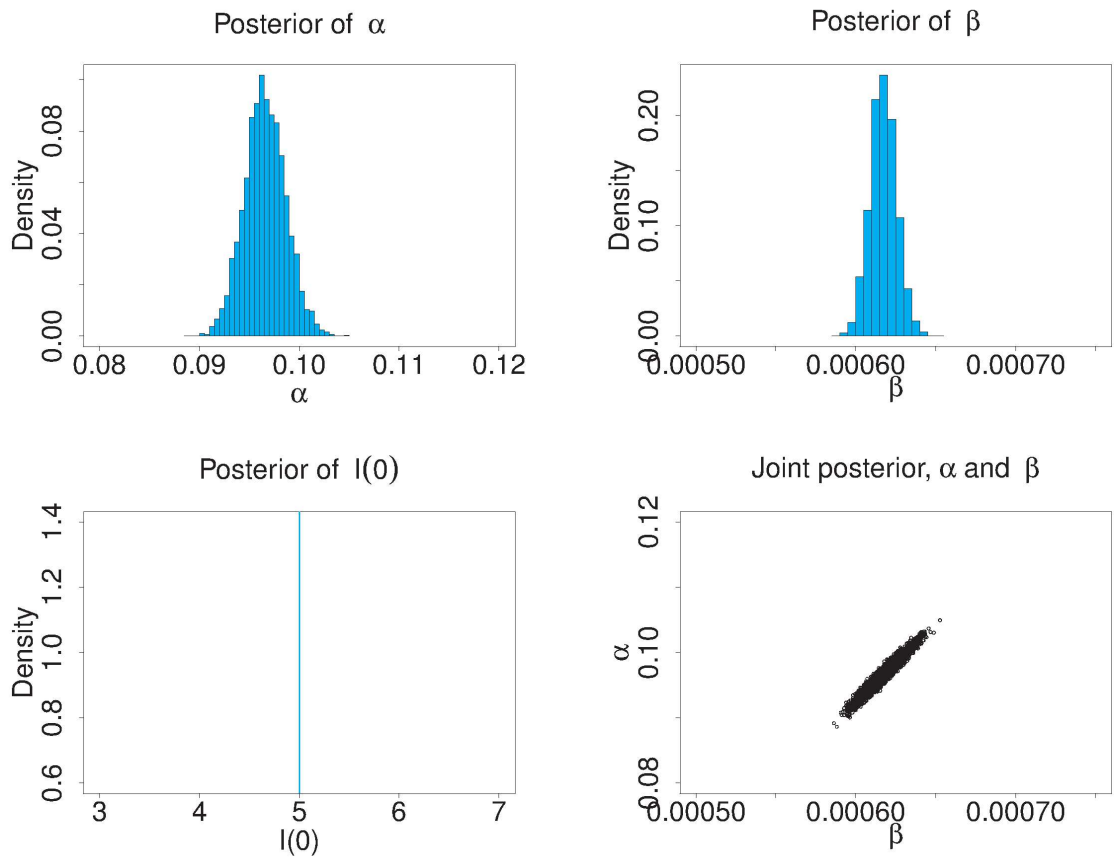}}
	\caption{The SIR-ODE model -- marginal and bivariate joint posterior distributions of sampled parameters  $\alpha,\beta \mbox{ and } I(0)$ obtained from the IMIS-Opt. The IMIS-Opt algorithm was run with $N_{0}=3000, D=3, B=1000, J=10000, N=1000$ (see Appendix \ref{app:ImplemIMIS-ShOpt} for implementation details). }
	\label{fig:pairsPlot_SIR_IMIS_opt}
\end{figure}

\begin{figure}[h]
	\centerline{\includegraphics[scale=0.34]{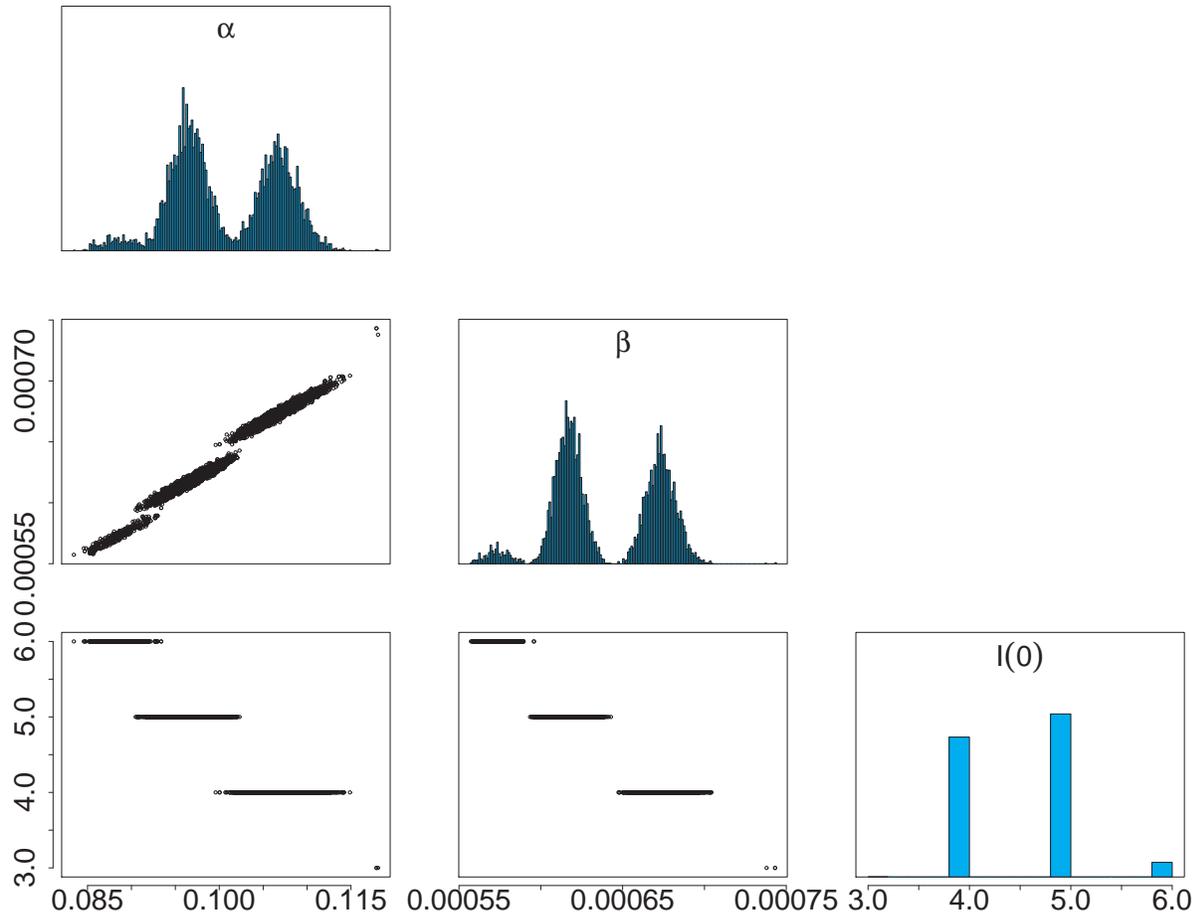}}
	\caption{The SIR-ODE model -- marginal (diagonal) and bivariate joint (off-diagonal) posterior distributions of sampled parameters  $\alpha,\beta \mbox{ and } I(0)$ obtained from the IMIS-ShOpt.   The IMIS-ShOpt algorithm was run with $N_{0}=3000, Q=10, D=3, B=1000, J=10000, N=1000$.}
	\label{fig:pairsPlot_SIR_IMIS_ShOpt}
\end{figure}

 Figure~\ref{fig:pairsPlot_SIR_IMIS_ShOpt} illustrates multi-modality and topological challenges of the posterior space of the SIR model. Marginal distributions of the two continuous parameters $\alpha$ and $\beta$ exhibit three isolated modes. Clouds in the bivariate plot of $\alpha$ and $\beta$ depicts the four modes corresponding to the discrete values of $I(0)=\{6,5,4,3\}$ from left to right. While the IMIS-ShOpt captures all the multiple modes in the posterior space, the IMIS-Opt gets trapped in the mode around the initial state $I(0) =5$ (Figure~ \ref{fig:pairsPlot_SIR_IMIS_opt}). The IMIS-Opt uses gradient optimization to optimize the target posterior distribution. Regardless of the many different starting points, the gradient optimization discovers only the global mode.

\section{Parameter estimation with IMIS-ShOpt using synthetic likelihood } \label{Sec:IMIS-ShOpt-SL}

In this section we introduce the IMIS-ShOpt with synthetic likelihood \citep{wood2010statistical}   which borrows ideas from the Approximate Bayesian Computation (ABC) framework. ABC methods \citep{Tavare505,pritchard1999population} provide a framework for inference in cases where the likelihood is intractable or very costly to evaluate,  but simulating data from the model is relatively easy.

In chaotic systems, likelihood-based inference breaks down because small changes in parameters induce big changes in the system states. To avoid the requirement of the tolerance levels and the distance measure needed in ABC, and to gain the efficiency from the Shotgun optimization  thereof, we
approximate the likelihood function with a synthetic likelihood \citep{wood2010statistical}.  The synthetic likelihood captures important dynamics in the data using the summary statistics.  Although synthetic likelihood approach employs ideas from the ABC framework, the log synthetic likelihood behaves like a conventional log likelihood in the limit, when the number of simulated data sets approaches infinity, but acts with reduced efficiency because of the lack of sufficient statistics.

Following \cite{wood2010statistical}, the synthetic likelihood can be constructed as follows. For  parameters $\boldsymbol{\theta}$, $N_{\boldsymbol{Z}}$ simulated data sets $\boldsymbol{Z}=\{\boldsymbol{Z}_{1},..,\boldsymbol{Z}_{N_{Z}}\}$ are generated from $P(\boldsymbol{Z} \mid \boldsymbol{\theta})$, and the vector of summary statistics $\boldsymbol{S(Z)}=\{\boldsymbol{s}(\boldsymbol{Z}_{1}),..,\boldsymbol{s}(\boldsymbol{Z}_{N_{\boldsymbol{Z}}})\}$ is calculated for each  simulated data set, exactly as the summary statistics $\boldsymbol{S(Y)}$ is calculated from the observed data. The mean of the $N_{\boldsymbol{Z}}$ summary statistics, $\hat{\boldsymbol{\mu}}_{\theta} = \frac{\sum\limits_{i=1}^{N_{\boldsymbol{Z}}} \boldsymbol{s}(\boldsymbol{Z}_{i})}{N_{\boldsymbol{Z}}}$, and the variance-covariance matrix,  $ \hat{\boldsymbol{\Sigma}}_{\boldsymbol{\theta}}$, are used to construct the synthetic likelihood as $MVN(\boldsymbol{S} \mid \hat{\boldsymbol{\mu}}_{\theta},  \hat{\boldsymbol{\Sigma}}_{\boldsymbol{\theta}})$, i.e.,

\begin{equation}
\mathcal{L}_{s}(\boldsymbol{\theta} \mid \boldsymbol{S(Y)})=-\frac{1}{2}(\boldsymbol{S(Y)}-\hat{\boldsymbol{\mu}}_{\theta})^{'}\hat{\boldsymbol{\Sigma}}_{\boldsymbol{\theta}}^{-1}(\boldsymbol{S(Y)}-\hat{\boldsymbol{\mu}}_{\theta}) -\frac{1}{2}log| \hat{\boldsymbol{\Sigma}}_{\boldsymbol{\theta}}|.  \label{eq:syntheticLogLik}
\end{equation}
When a set of candidate summary statistics is available, the target likelihood is defined over the entire set of available summary statistics.

The objective function in the Shotgun optimization step uses different approximations to the synthetic likelihood $\mathcal{L}_{s}(\boldsymbol{\theta} \mid \boldsymbol{S(Y)})$ defined over subsets of the entire set of summary statistics. These approximations to the target synthetic likelihood   might explore different regions of the posterior space. The strategy of defining different approximations to the synthetic likelihood was used to construct several different optimization criteria in the Shotgun optimization stage of the IMIS-ShOpt, one for each random subset of summary statistics. The Hessian matrix is calculated using the target synthetic likelihood which operates on the entire set of the available summary statistics.

The proposed IMIS-ShOpt algorithm draws samples from the posterior distribution of the parameters of interest in models where likelihood function is computationally very costly to evaluate. In the initial stage, $N_{0}$ samples $\{ \boldsymbol{\theta}_{1},\boldsymbol{\theta}_{2},...,\boldsymbol{\theta}_{N_{0}}\}$ are drawn from the prior distribution $P(\boldsymbol{\theta})$. Sampling weights are calculated using the target synthetic likelihood $\mathcal{L}_{s}(\boldsymbol{\theta} \mid \boldsymbol{S(Y)})$, defined over the entire set of available summary statistics.

The pseudo-code of the IMIS-ShOpt algorithm with synthetic likelihood is given in the Algorithm \ref{alg:IMIS-ShOpt-SL}.

\begin{algorithm}[]  
	\caption{The IMIS-ShOpt with synthetic likelihood}
	\textbf{Goal: Parameter estimation} \\
	\textbf{Input:} 	Data, likelihood function, synthetic likelihood function, prior distribution and the model.\\
	Initialize $N$ -- the number of iterations, $B$ --the number of incremental points, $D$ -- the number of different initial points for the optimization, $Q$ -- the number of different optimization criteria, $N_{0}$ -- the number of initial samples from the prior and $J$ -- the number of re-sampled points. \\
	\textbf{Initial stage:} 
	Draw $N_{0}$ samples $\boldsymbol{\Theta}_{0}=\{ \boldsymbol{\theta_{1}},\boldsymbol{\theta_{2}},...,\boldsymbol{\theta_{N_{0}}}\}$ from the prior distribution $P(\boldsymbol{\theta})$.
	\label{alg:IMIS-ShOpt-SL}
	\begin{algorithmic}
		\For{$k=1:N$}	
		\If{$k=1$} 
		\State For each $\boldsymbol{\theta}_{i}$, $i=1,..,N_{0}$,	simulate ${N_{\boldsymbol{Z}}}$ vectors of replicate data $\boldsymbol{Z}_{i}=\{\boldsymbol{Z}_{1},..,\boldsymbol{Z}_{N_{\boldsymbol{Z}}}\}$ from the model, $P(\boldsymbol{Z} \mid \boldsymbol{\theta}_{i})$.
		\State For each $\boldsymbol{\theta}_{i}$, $i=1,..,N_{0}$, calculate the vector of entire set of available summary statistics, $\boldsymbol{S(Z)}=\{\boldsymbol{s}(\boldsymbol{Z}_{1}),..,\boldsymbol{s}(\boldsymbol{Z}_{N_{\boldsymbol{Z}}})\}$ and construct the synthetic likelihood using (\ref{eq:syntheticLogLik}). 
		\State For each $\boldsymbol{\theta}_{i}, i=1,..,N_{0}$ calculate the sampling weights, 
		\begin{equation}
		w_{i}^{(k)}=\frac{\mathcal{L}_{s}(\boldsymbol{\theta}_{i} \mid \boldsymbol{S(Y)})}{\sum\limits_{j=1}^{N_{0}} \mathcal{L}_{s}(\boldsymbol{\theta}_{j} \mid \boldsymbol{S(Y)})}
		\end{equation}
		\State \textbf{Optimization stage:}	
		\For{$d = 1: D$}
		\State Find the d-th maximum weight point
		$\boldsymbol{\theta}_{d}^{(initial)}=\underset{\boldsymbol{\theta}}{\operatorname{argmax}} \mbox{ } \boldsymbol{w}^{(k)} (\boldsymbol{\theta})$, $\boldsymbol{\theta} \in \boldsymbol{\Theta}_{d-1}$ to initialize Q optimizers. 			
		\For{$q = 1: Q$}
		\State  Use q-th optimization method to optimize $\boldsymbol{\theta}$, the objective function is $\mathcal{L}_{s}(\boldsymbol{\theta} \mid \boldsymbol{S(Y)})$ based on a subset of summary statistics, i.e., obtain local  maxima $\boldsymbol{\theta}_{d,q}^{(Opt)}$. Obtain the corresponding inverse Hessian $\boldsymbol{\Sigma}_{d,q}^{(Opt)}$ using the target synthetic likelihood.
		\State Update $\boldsymbol{\Theta}_{d}$ by excluding $\frac{N_{0}}{DQ}$ nearest neighbor points, $\boldsymbol{\theta}_{k} \in \boldsymbol{\Theta}_{d-1}$, that minimize the Mahalanobis distance,
		\begin{equation}
		( \boldsymbol{\theta}_{k}-\boldsymbol{\theta}_{d,q}^{(Opt)})^{'} (\boldsymbol{\Sigma}_{d,q}^{(Opt)})^{-1} (\boldsymbol{\theta}_{k}-\boldsymbol{\theta}_{d,q}^{(Opt)} ). \end{equation}
	
		\algstore{alg:IMIS-opt1}
	\end{algorithmic}
\end{algorithm}
\begin{algorithm}[H]	
	\caption*{\textbf{Algorithm \ref{alg:IMIS-ShOpt-SL}} The IMIS-ShOpt with synthetic likelihood - continued}
	
	\begin{algorithmic}
		\algrestore{alg:IMIS-opt1}
        \State Draw $B$ samples $\boldsymbol{\theta}_{1:B} \sim MVN(\boldsymbol{\theta}_{d,q}^{(Opt)},\boldsymbol{\Sigma}_{d,q}^{(Opt)})$; add these points to the importance sampling distribution $P(\boldsymbol{\theta} \mid \boldsymbol{Y})$ and
		evaluate $H_{k}=MVN(\boldsymbol{\theta}_{1:B} \mid \boldsymbol{\theta}_{d,q}^{(Opt)},\boldsymbol{\Sigma}_{d,q}^{(Opt)})$.
		\EndFor
		\EndFor
		\Else

        		\State \textbf{Importance sampling stage:}	
		\State For each $\boldsymbol{\theta}_{i}, i=1,..,N_{k}$ calculate weights:	
		\begin{equation}
		w_{i}^{(k)}=\frac{c P(\boldsymbol{\theta}_{i}) \mathcal{L}_{s}(\boldsymbol{\theta}_{i} \mid \boldsymbol{S(Y)})   }{ \frac{N_{0}}{N_{k}}P(\boldsymbol{\theta}_{i}) +\frac{B}{N_{k}}\sum\limits_{s=1}^{k} H_{s}(\boldsymbol{\theta}_{i}) },
		\end{equation}
		\State where $N_{k}=N_{0}+B(QD+k)$ and $c=1/\sum\limits_{i=1}^{N_{k}} w_{i}^{(k)}$.
		\State  Choose a maximum weight input, $\boldsymbol{\theta}_{k}$, and estimate $\boldsymbol{\Sigma}_{k}$ as the weighted covariance of B inputs with smallest Mahalanobis distance,
		\[w_{p}(\boldsymbol{\theta})\left(\boldsymbol{\theta}-\boldsymbol{\theta}_{k}\right)^{'}(\boldsymbol{\Sigma_{\pi}})^{-1}\left(\boldsymbol{\theta}-\boldsymbol{\theta}_{k}\right), \]
		where the weights are $w_p(\boldsymbol{\theta})=c_{1}(\boldsymbol{w}^{(k)}+1/N_{k})$, $\Sigma_{\pi}$ is the covariance of the initial importance distribution and $c_{1}=1/w_p(\boldsymbol{\theta})$.
		\State Draw $B$ samples $\boldsymbol{\theta}_{1:B} \sim MVN(\boldsymbol{\theta}_{k},\boldsymbol{\Sigma}_{k})$; add these points to the importance sampling distribution and 
		evaluate $H_{k}=MVN(\boldsymbol{\theta}_{1:B} \mid \boldsymbol{\theta}_{m,k},\boldsymbol{\Sigma}_{k})$.
		\EndIf 
		\If  {$\sum\limits_{1}^{N_{k}}(1-(1-w^{(k)})^{J}) \geq J(1-\exp{(-1)})$ i.e., importance sampling weights are approximately uniform}  exit for loop 
		\EndIf
		\EndFor
		\State \textbf{Re-sampling stage:}	
		\State Re-sample $J$ points with replacement from $\{\boldsymbol{\theta_{1}},..,\boldsymbol{\theta_{N_{k}}}\}$ and weights $w^{(k)}$.
		
	\end{algorithmic}
\end{algorithm}

\subsection{Illustration of the IMIS-ShOpt with synthetic likelihood}

Consider a chaotic stochastic difference model, where full likelihood-based inference fails.  The model exhibits intractable or expensive-to-evaluate likelihoods, but it is relatively easy to simulate data from the  model.

Following \cite{gilpin1973global}, the ecological theta-Ricker model, states that the abundance of the population in the next time point, $N_{t+1}$, is equal to the abundance at the current time point $N_{t}$, multiplied by the exponent of the growth rate, $\exp{\left(r(1-\frac{N_{t}}{K})^{\tilde{\theta}}+\epsilon_{t}\right)}$, over the time step $t$.  The process noise, also known as environmental noise is modeled as  $\epsilon_{t} \sim N(0,\sigma_{p}^{2})$ and $K$ quantifies carrying capacity. The theta-Ricker model can be written as follows, 
\begin{equation}
N_{t+1}=N_{t}\exp{ \left( r\left(1- \left(\frac{N_{t}}{K}\right)^{\tilde{\theta}}\right) +\epsilon_{t}\right)}, 
\end{equation}.

The theta-Ricker model is defined with parameters $\boldsymbol{\theta}=(r,\phi,\sigma_{p}^{2},\tilde{\theta})^{'}$. The data are outcomes of the Poisson distribution with mean $\phi N_{t}$, where $\phi$ is a scaling parameter, 
\begin{equation*}
y_{t} \sim Poisson(\phi N_{t}).
\end{equation*}

The IMIS-ShOpt algorithm was used to estimate the parameters of the theta-Ricker model. The data were simulated from  $\boldsymbol{\theta}=(\log{r}=0.5,\phi=4,\sigma^{2}=0.01,\log{\tilde{\theta}}=1)^{'}$ at $T$=50 time steps with initial population $N_{0}=3$ and  $K=100$. Prior distributions were defined independently,  $\log{r} \sim N(0.5,1),  \phi \sim \chi^2(df=4),  \sigma_{p}^{2} \sim IGamma(shape=2,scale=0.05), \log{\tilde{\theta}}=N(1,1)$. The IMIS-ShOpt was initialized with $B=1000,J=3000,N_{0}=10000,N=500, D=10,Q=3,N_{Z}=30$.

The set of summary statistics used in IMIS-ShOpt is a modification of the set from \cite{golchi2016sequentially},
\begin{eqnarray}
\boldsymbol{S(Y)}= \{ median(\boldsymbol{Y}), \sum\limits_{i=1}^{n}\frac{y_{i}}{n}, \frac{\sum\limits_{i=1}^{n}y \mathbb{I}_{(1,\infty)}(y_{i})}{\sum\limits_{i=1}^{n} \mathbb{I}_{(1,\infty)}(y_{i})},  \sum\limits_{i=1}^{n}y \mathbb{I}_{(10,\infty)}(y_{i}),\sum\limits_{i=1}^{n} \mathbb{I}_{0}(y_{i}), \nonumber \\ Quantile_{0.75}(\boldsymbol{Y}),max(\boldsymbol{Y}),\sum\limits_{i=1}^{n} \mathbb{I}_{(100,\infty)}(y_{i}),\sum\limits_{i=1}^{n} \mathbb{I}_{(300,\infty)}(y_{i}), \nonumber \\ \sum\limits_{i=1}^{n} \mathbb{I}_{(500,\infty)}(y_{i}),  \sum\limits_{i=1}^{n}y \mathbb{I}_{(800,\infty)}(y_{i}) \}.    \label{eq:summaryStats}
\end{eqnarray}

The target optimization function is the synthetic likelihood in (\ref{eq:syntheticLogLik}) defined over the entire set of summary statistics $\boldsymbol{S(Y)}$ in (\ref{eq:summaryStats}).
The $q$-th optimization method in the Shotgun optimization strategy initialized at fixed $d$, corresponds to an approximation to the target synthetic likelihood, $\mathcal{L}_{s}(\boldsymbol{\theta} \mid \boldsymbol{\tilde{S}(Y)})$, defined over a random subset of seven unique summary statistics from the entire set of summary statistics, $\boldsymbol{\tilde{S}(Y)}= \{ s_{i},s_{j},s_{k},s_{l},s_{m},s_{o} ,s_{p}\mid i,j,k,l,m,o,p = 1,..,11\} \subseteq
\boldsymbol{S(Y)}$. 
Different approximations to the target synthetic likelihood explore different parts of the posterior space, and therefore, combining the results should lead to discovering all important posterior modes. Hence, multiple optimization criteria in the Shotgun optimization were defined to correspond to distinct approximations to the synthetic likelihood.

Although the locations of the posterior modes were discovered using different approximations to the target synthetic likelihood, the Hessian matrices of the posterior modes were obtained numerically using the target synthetic likelihood, $\mathcal{L}_{s}(\boldsymbol{\theta} \mid \boldsymbol{S(Y)})$.

\subsection{Results}
\begin{figure}[ht]
	\centerline{\includegraphics[scale=0.28]{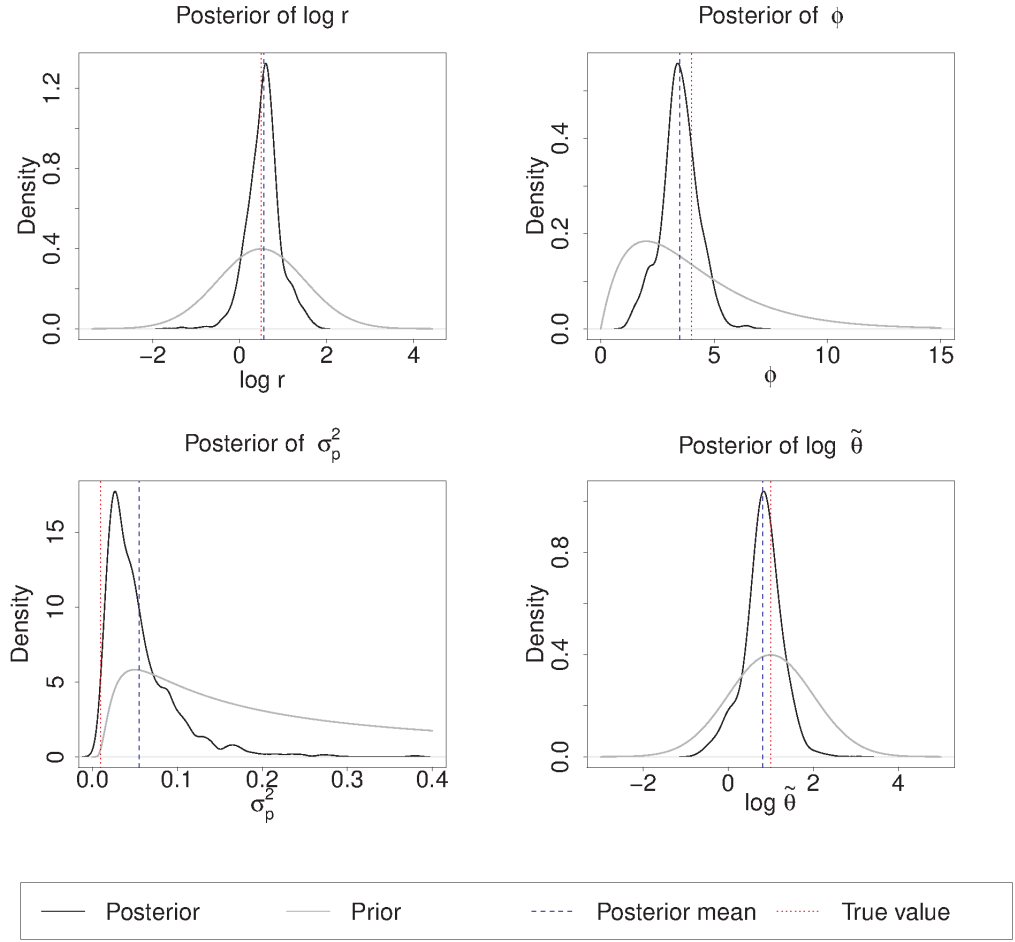}}
	\caption{The theta-Ricker model -- marginal posterior distributions of the parameters obtained from the final re-sampling stage. The vertical lines are drawn at the posterior mean (blue dashed) and the true value (red dotted). The gray distributions represent the priors.}
	\label{fig:thetaRickerPosterior}
\end{figure}

\begin{figure}[ht]
	\centerline{\includegraphics[scale=0.28]{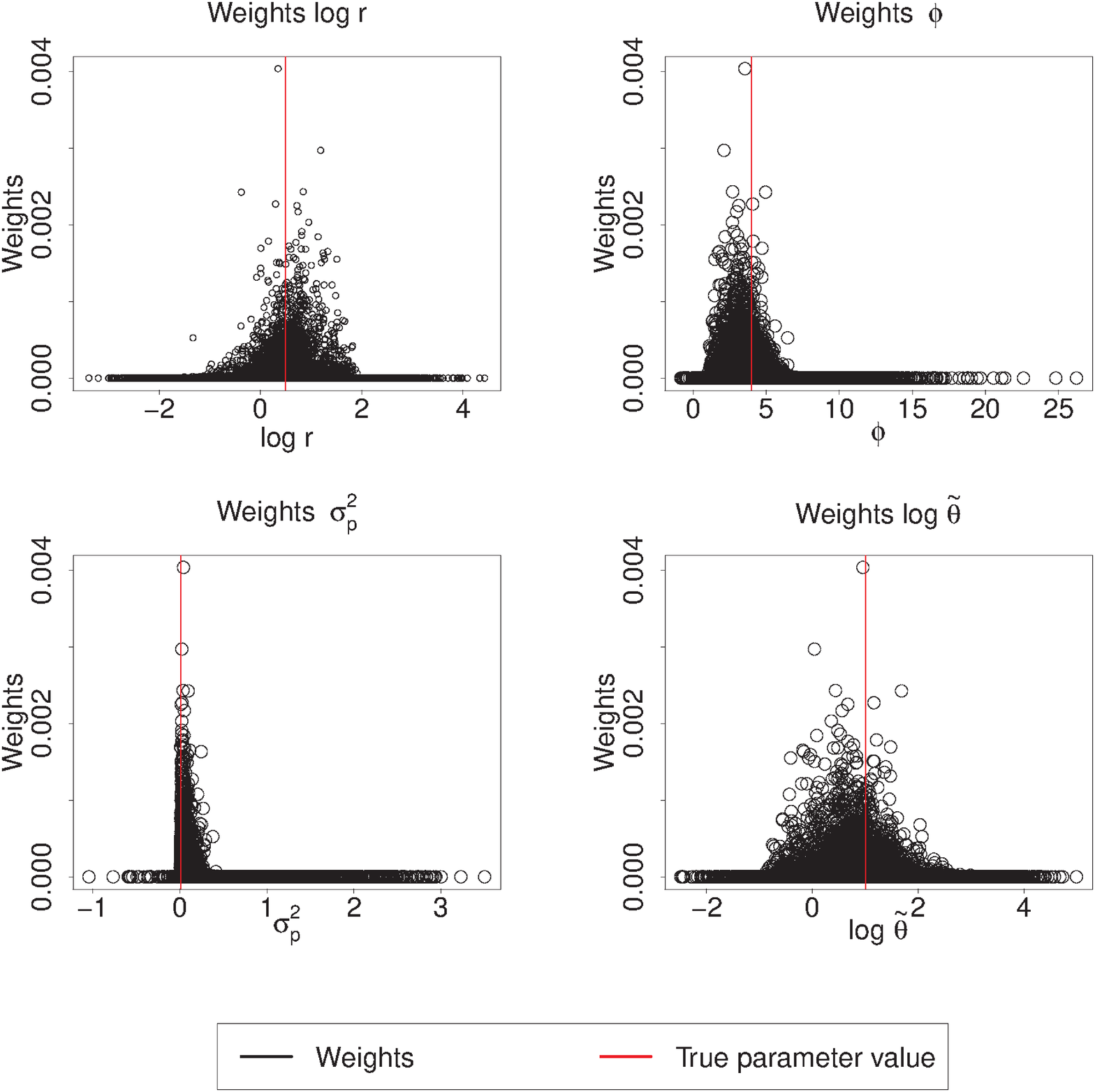}}
	\caption{The theta-Ricker model -- weights of the particles in the importance sampling distribution before re-sampling. The vertical lines are drawn at the true parameter values.}
	\label{fig:thetaRickerWeights}
\end{figure}

The IMIS-ShOpt with synthetic likelihood produces reasonable parameter estimates. The results, presented as kernel density estimates of the approximate marginal posteriors, are given in  Figure~\ref{fig:thetaRickerPosterior}. Figure~\ref{fig:thetaRickerWeights} shows that the weights of all the particles in the importance sampling distribution before the final re-sampling stage are non-zero in the neighborhood of the true parameter values. In addition, Figure~\ref{fig:thetaRickerWeights} demonstrates that  before the final re-sampling stage the importance sampling distribution  of the process noise variance, $\sigma_{p}^{2}$,  contains particles with negative values. 
These points are added to the importance sampling distribution during the optimization and importance sampling stage, but do not survive the final re-sampling stage because they have zero weights as shown in Figures~\ref{fig:thetaRickerPosterior} and \ref{fig:thetaRickerWeights}. Rather than harming the importance sampling distribution, the negative-valued points help in better exploration of the posterior surface.

The Shotgun optimization helps exploring the parameter space through the approximations to the target synthetic likelihood.
Namely, the target synthetic likelihood, which employs the entire set of the summary statistics, exhibits narrow spiky modes which leads to optimization difficulties. Approximations to the target synthetic likelihood constructed by randomly chosen subsets of seven summary statistics, are more diffuse then the target synthetic likelihood, and hence, easier to optimize. 
Shotgun optimization combines results from different approximations to the target synthetic likelihood, thus resulting in more fully exploration of the parameter space.

\section{Discussion} \label{sec:Conclusions_ShOpt}

This paper proposes a general optimization framework, the Shotgun optimization, which relies on the idea that no single method outperforms other methods in every situation. Different methods employ different model variations which leads to exploring different regions of the posterior space. Combining the results from different methods balances discovery of global and local modes, which results in more fully explored posterior space. Some methods produce better estimates of the parameters, while others introduce bias. Merging the results from different methods together can overcome the introduced bias.

The Shotgun optimization strategy is a general framework which can be applied in any model type. Given a model type,  competing parameter estimation methods deal with the posterior topologies in different ways, which leads to exploring diverse and potentially informative locations of the parameter space. The Shotgun optimization incorporates results from different competing methods or from different optimization criteria, and ensures that the parameter space is more fully explored. In addition, the Shotgun optimization is computationally efficient, since it can be easily parallelized. For instance,  in the FhN-ODE model, the Shotgun optimization method runs in parallel  the following three parameter estimation techniques for ODE models: the Non-linear Least Squares, the Two-Stage and the Generalized Profiling. Each of the methods discovers either a local mode or the global mode, but combined together the three methods find all the important modes. Similarly, in the SIR-ODE model, the Shotgun optimization consists of fitting the Non-linear Least Squares locally at several possible locations of the posterior modes corresponding to different values of the initial infection state. The Shotgun optimization strategy in the IMIS-ShOpt with synthetic likelihood merges results from different optimization criteria defined by different approximations to the target synthetic likelihood. Each approximation to the target synthetic likelihood corresponds to a randomly chosen subset of summary statistics from the entire collection of seven summary statistics thus exploring different locations of the posterior space.

\bigskip
\begin{center}
{\large\bf SUPPLEMENTAL MATERIALS}
\end{center}
R code of the implemented examples is stored in a zipped archive (codeSubmit.zip).



\section{Acknowledgments}

This work was supported by grant from Natural Sciences and Engineering Research Council of Canada (RGPIN04040-2014) and the Department of Statistics and Actuarial Sciences at Simon Fraser University. 
The authors would like to thank Dr. Luke Bornn, Dr. Derek Bingham, Dr. Liangliang Wang and Dr. Russell Steele for the constructive discussions, and Dr. Michael Jack Davis for proof reading the manuscript.

\bibliographystyle{apa}
\bibliography{IMIS_ShOpt}

\section{Appendices}

	\subsection{Implementation of IMIS-Opt and IMIS-ShOpt, SIR model} \label{app:ImplemIMIS-ShOpt}

The IMIS-ShOpt for the SIR model draws samples from the target posterior $P(\alpha,\beta,I(0) \mid \boldsymbol{Y})$ by sampling the two continuous parameters $\alpha, \beta$ conditionally on the $I(0)$ while updating $I(0)$ uniformly over $\{1,2..,10\}$.   The algorithm starts with initial particles $\{\alpha,\beta,I(0)\}$ from the prior given in (\ref{eq:priors}), and then calculates initial weights 
using the likelihood in (\ref{eq:llik_ShOpt}). The Shotgun optimization, optimizes the sum of squared error function in (\ref{eq:NLSOptim}), by finding local maxima of $\alpha ,\beta$ conditional on $I(0) \in \{1,2,..,10\}$, i.e., $(\alpha, \beta \mid I(0) \in \{1,2,..,10\})$. For each newly discovered local mode, B samples $(\alpha,\beta)$ are drawn from the multivariate Gaussian, while $I(0)$ is updated with the corresponding value from $\{1,2,..,10\}$. Similarly, in the importance sampling stage, the maximum weight point is selected and the weighted covariance is calculated using the $\alpha, \beta \mid I(0) \in \{1,2,..,10\}$. The new B samples $(\alpha, \beta)$ are drawn from the multivariate Gaussian, while fixing the B samples from $I(0)$ to the value of $I(0)$ from the currently selected maximum weight point.
Pseudo code of the implementation of the IMIS-ShOpt on the SIR model is given in the Algorithm~\ref{alg:IMIS-ShOpt_SIR}.

Similar to the IMIS-ShOpt, the IMIS-Opt on the SIR model,
updates the two continuous parameters $\alpha, \beta$ conditionally on the $I(0)$ while updating $I(0)$ uniformly over $\{1,2..,10\}$.
The optimization stage is implemented as follows. Optimization of   $\boldsymbol{\theta}=\left(\alpha,\beta, I(0) \right)^{'}$,  was carried out by first optimizing the conditional posterior distribution $P(\alpha,\beta \mid  I(0), \boldsymbol{Y}, \tau)$  for each $I(0)=\{1,..,10\}$. Then, out of the ten optimized values $\left(\alpha_{max},\beta_{max},I(0) \mid I(0) \  \in \{1,2,..,10\}\right)^{'}$, the one that maximizes the posterior distribution $P(\alpha,\beta,I(0) \mid  \boldsymbol{Y}, \tau)$ was chosen as a maximum. Hence, instead of keeping all the 10 optima and using them to repopulate the importance sampling distribution as in IMIS-ShOpt, the IMIS-Opt uses only one optima to repopulate the importance sampling distribution. The Hessian matrix was obtained using the conditional posterior $P(\alpha,\beta \mid  I(0), \boldsymbol{Y}, \tau)$  for the corresponding $I(0)=\{1,..,10\}$. 
The importance sampling stage follows the importance sampling stage in the IMIS-ShOpt.

\begin{algorithm}[H]  
	\caption{IMIS-ShOpt for the SIR model}
	\textbf{Goal: Draw samples from the target distribution $P(\boldsymbol{\theta} \mid \boldsymbol{Y})$ of the SIR model, where $\boldsymbol{\theta} =(\alpha,\beta, I(0))^{'}$}.
	
	\textbf{Input:} Data, model, likelihood function, prior distribution, $B$ - the number of incremental points, $D$ - the number of different initial points for the optimization, $N_{0}$ - the number of the initial samples from the prior and $J$ - the number of re-sampled points, $N$ - the number of iterations.
	
	\textbf{Initial stage:}  Draw $N_{0}$ samples $\boldsymbol{\Theta}_{0}=\{ \boldsymbol{\theta_{1}},\boldsymbol{\theta_{2}},...,\boldsymbol{\theta_{N_{0}}}\}$ from the prior distribution $P(\boldsymbol{\theta})$ as per (\ref{eq:priors}).
	\label{alg:IMIS-ShOpt_SIR}
	\begin{algorithmic}
		
		\For{$k=1:N$}	
		
		\If{k=1} 
		\State For each $\{\boldsymbol{\theta_{i}}, i=1,..,N_{0}\}$ calculate the sampling weights:	
		\begin{equation}
		w_{i}^{(1)}=\frac{P\left(\boldsymbol{Y} \mid \boldsymbol{\theta_{i}}\right)}{\sum\limits_{j=1}^{N_{0}}{P(\boldsymbol{Y} \mid \boldsymbol{\theta_{j}})}},
		\end{equation}
		using the likelihood function  in (\ref{eq:llik_ShOpt})
		\State \textbf{Optimization stage:}	
		\For{$d = 1: D$}
		
		\State Find the d-th maximum weight point
		$\boldsymbol{\theta}_{d}^{(initial)}=\underset{\boldsymbol{\theta}}{\operatorname{argmax}} \mbox{ } \boldsymbol{w}^{(k)} (\boldsymbol{\theta})$, $\boldsymbol{\theta} \in \boldsymbol{\Theta}_{d-1}$ to initialize $Q$ optimizers. 
		
		\For{$q = 1: 10$}
		\State Let $\check{\boldsymbol{\theta}}= (\boldsymbol{\theta} \mid I(0)=q)$ denote a vector of parameters of interest conditional on $I(0)=q$.
		\State  Use NLS method as per (\ref{eq:NLSOptim}) initialized at $\boldsymbol{\theta}_{d}^{(initial)}$ to obtain local  maxima
		\begin{equation}
		\check{\boldsymbol{\theta}}_{d,q}^{(Opt)}=\arg\min_{\check{\boldsymbol{\theta}}} \sum\limits_{s=1}^{S}
		\sum\limits_{j=1}^{n_{s}} \left[y_{sj}-\boldsymbol{X}(\boldsymbol{\theta},t_{sj})\right]^{2},
		\end{equation}
		and obtain the corresponding inverse negative Hessian, $\boldsymbol{\Sigma}_{d,q}^{(Opt)}$, using the conditional target posterior $P(\alpha,\beta \mid I(0)=q,\boldsymbol{Y} )$.
		\State Update $\boldsymbol{\Theta}_{d}$ by excluding $\frac{N_{0}}{QD}$ nearest neighbor points, $\boldsymbol{\theta}_{k} \in \boldsymbol{\Theta}_{d-1}$, that minimize the Mahalanobis distance,

		\begin{equation}
		(\check{\theta}_{k} - \check{\boldsymbol{\theta}}_{d,q}^{(Opt)}  )^{'} (\boldsymbol{\Sigma}_{d,q}^{(Opt)})^{-1} (\check{\theta}_{k} -\check{\boldsymbol{\theta}}_{d,q}^{(Opt)} ).
		\end{equation}
		\algstore{alg:IMIS-ShOptSIR_interupt}
	\end{algorithmic}
\end{algorithm}
\begin{algorithm}[H]	
	\caption*{\textbf{Algorithm \ref{alg:IMIS-ShOpt_SIR}} IMIS-ShOpt for the SIR model - continued }
	\begin{algorithmic}
		\algrestore{alg:IMIS-ShOptSIR_interupt}
        	\State Draw B samples $\check{\boldsymbol{\theta}}_{1:B} \sim MVN(\check{\boldsymbol{\theta}}_{d,q}^{(Opt)},\boldsymbol{\Sigma}_{d,q}^{(Opt)})$ and repopulate B samples with $I(0)=q$; add these points to the importance sampling distribution and
		evaluate $H_{k}=MVN(\check{\boldsymbol{\theta}}_{1:B}  \mid \check{\boldsymbol{\theta}}_{d,q}^{(Opt)} ,\boldsymbol{\Sigma}_{d,q}^{(Opt)})$.
		\EndFor
		\EndFor
		
		\Else
		
		\State \textbf{Importance sampling stage:}	
		\State For each $\{\boldsymbol{\theta_{i}}, i=1,..,N_{k}\}$ calculate weights,
		\State
		\begin{equation}
		w_{i}^{(k)}=\frac{cP(\boldsymbol{Y} \mid \boldsymbol{\theta_{i}})P(\check{\boldsymbol{\theta}}_{i})}{ \frac{N_{0}}{N_{k}}P(\check{\boldsymbol{\theta}}_{i}) +\frac{B}{N_{k}}\sum\limits_{s=1}^{k} H_{s}(\check{\boldsymbol{\theta}}_{i}) },
		\end{equation}
		\State where $N_{k}=N_{0}+B(D+k)$ and $c=1/\sum\limits_{i=1}^{N_{k}} w_{i}^{(k)}$ is the normalizing constant.
		\State  Choose the maximum weight input $\boldsymbol{\theta}_{k}$ and extract $s=I(0)$ for this point; then estimate $\boldsymbol{\Sigma}_{k}$ as the weighted covariance of B inputs with smallest Mahalanobis distance,

		\[w_p(\boldsymbol{\theta})\left( \check{\boldsymbol{\theta}}-\check{\boldsymbol{\theta}}_{k}\right)^{'}(\boldsymbol{\Sigma_{\pi}})^{-1}\left( \check{\boldsymbol{\theta}}-\check{\boldsymbol{\theta}}_{k} \right),\]
		where 
		$\check{\boldsymbol{\theta}}= (\boldsymbol{\theta}_{k} \mid I(0)=s)$ corresponds to the vector of parameters of interest conditional on the current value of $I(0)=s$,        	
		the weights $w_p(\boldsymbol{\theta})$ are proportional to the average of the importance weights and the uniform weights $\frac{1}{N_{k}}$, ${\boldsymbol{\Sigma}_{\pi}}$ is the covariance of the initial importance distribution.
		\State Draw B samples $\check{\boldsymbol{\theta}}_{1:B} \sim MVN( \check{\boldsymbol{\theta}}_{k},\boldsymbol{\Sigma}_{k})$; add these points to the importance sampling distribution  and re-populate the new B samples for $I(0)=s$; then
		evaluate $H_{k}=MVN( \check{\boldsymbol{\theta}}_{1:B} \mid \check{\boldsymbol{\theta}}_{k},\boldsymbol{\Sigma}_{k})$.
		\EndIf 
		\If  {$\sum\limits_{1}^{N_{k}}(1-(1-w^{(k)})^{J}) \geq J(1-\exp{(-1)})$ i.e., importance sampling weights are approximately uniform}  exit for loop 
		
		\EndIf
		\EndFor
		\State \textbf{Re-sampling stage:}	
		
		\State Re-sample J points with replacement from $\{\boldsymbol{\theta}_{1},..,\boldsymbol{\theta}_{N_{k}}\}$ and weights $(w_{1},..,w_{N_{k}})^{'}$.
	\end{algorithmic}
\end{algorithm}

Both algorithms, the IMIS-Opt and the IMIS-ShOpt, used diffuse prior densities for the SIR-ODE model parameters. As a result, a big proportion of the initial importance samples drawn from the prior distribution fall outside the domain of the ODE model where the solution does not exist.
For algorithmic convenience, the  log-likelihood for the points outside the domain of the ODE system was set to take very small values (e.g., - 999999) so that the weights of these points were effectively zero. Hence, both algorithms were initialized with only few non-zero weight samples from the prior. The IMIS-ShOpt employed Q=10 optimization methods initialized at the highest weight point to discover 10 different modes, whereas the IMIS-Opt used only one optimization routine initialized at
the highest weight point to find only one mode. The rest of the non-zero weight initial points were within the basin of attraction of the previously discovered  modes, and hence, they were excluded from the set of candidates initial optimization points.

Optimization step in both algorithms continued initializing the optimizers with zero-weight points, which in turn did not contribute in discovering new modes. These 'bad' points could have been either physically removed from the importance distribution or kept with their likelihood set to an extremely small value (e.g, -9999999). Keeping the 'bad' points in the importance sampling distribution did not harm the convergence of the both algorithms, because ultimately the highest-weight points were re-sampled in the final stage.









\end{document}